\documentclass[AMA,STIX1COL]{WileyNJD-v2}
\articletype{}%
\received{XX.XX.XXXX}
\revised{YY.YY.YYYY}
\accepted{ZZ.ZZ.ZZZZ}

\usepackage{booktabs}  
\usepackage{rotating}  
\usepackage{graphicx}  
\usepackage{amsmath}

\providecommand{\binomdistn}{\mathrm{Binomial}}
\providecommand{\normaldistn}{\mathrm{Normal}}

\DeclareMathOperator{\logit}{logit}

\providecommand{\Efifty}{E_{50}}        
\providecommand{\Emax}{E_\mathrm{max}}  
\DeclareMathOperator{\emax}{emax}       

\begin{document}

  \title{A straightforward meta-analysis approach for oncology phase~I dose-finding studies}
    
  \author[1]{Christian R\"{o}ver}
  \author[2,3,4]{Moreno Ursino}
  \author[1,5]{Tim Friede}
  \author[3,4,5]{Sarah Zohar}
  \authormark{C.~R\"{O}VER, M.~URSINO, T.~FRIEDE, S.~ZOHAR}

  \address[1]{\orgdiv{Department of Medical Statistics}, \orgname{University Medical Center G\"{o}ttingen}, \orgaddress{\state{G\"{o}ttingen}, \country{Germany}}}
  \address[2]{\orgdiv{Unit of Clinical Epidemiology}, \orgname{AP-HP, CHU Robert Debré, Université de Paris, Inserm CIC-EC 1426}, \orgaddress{\state{Paris}, \country{France}}}
  \address[3]{\orgdiv{Inserm }, \orgname{Centre de Recherche des Cordeliers, Université de Paris, Sorbonne Université }, \orgaddress{\state{Paris}, \country{France}}}
  \address[4]{\orgdiv{HeKA}, \orgname{Inria Paris}, \orgaddress{\state{Paris}, \country{France}}}
  \address[5]{Co-last authors as seniors}
  \corres{*Christian R\"{o}ver, \email{christian.roever@med.uni-goettingen.de}}

\abstract[Summary]{
  Phase~I early-phase clinical studies aim at investigating the safety and the underlying \emph{dose-toxicity} relationship of a drug or combination. While little may still be known about the compound's properties, it is crucial to consider quantitative information available from any studies that may have been conducted previously on the same drug. A meta-analytic approach has the advantages of being able to properly account for between-study heterogeneity, and it may be readily extended to prediction or shrinkage applications. Here we propose a simple and robust two-stage approach for the estimation of maximum tolerated dose(s) (MTDs) utilizing penalized logistic regression and Bayesian random-effects meta-analysis methodology.  Implementation is facilitated using standard \textsf{R}~packages. The properties of the proposed methods are investigated in Monte-Carlo simulations. The investigations are motivated and illustrated by two examples from oncology.
}

\jnlcitation{\cname{\author{C. R\"{o}ver}, 
                    \author{M. Ursino},
                    \author{T. Friede}, and
                    \author{S. Zohar}} 
             (\cyear{2021}), 
             \ctitle{A simple meta-analysis approach for exploratory phase~I dose-finding studies}, 
             \cjournal{Statistics in Medicine}, \cvol{2021;00:0--0}.}

\maketitle

\footnotetext{\textbf{Abbreviations:} 
  BLRM, Bayesian logistic regression model;
  CI, credible interval; 
  CRM, coninual reassessment method; 
  DLT, dose-limiting toxicity; 
  FLAC, Firth's logistic regression with added covariate; 
  GLM, generalized linear model; 
  MAC, meta-analytic-combined; 
  MAP, meta-analytic-predictive; 
  MED, minimum effective dose;
  ML, maximum likelihood; 
  MTD, maximum tolerated dose; 
  NNHM, normal-normal hierarchical model;
  OR, odds ratio}

\section{Introduction}\label{sec:intro}
  Phase~I dose-finding studies are the first-in-human studies of the clinical development which aim at estimating the maximum tolerated dose (MTD) of a drug or a combination of molecules.  The MTD is usually defined as the dose level associated with a probability of occurrence of treatment-related adverse events, so-called dose-limiting toxicities (DLTs), at a pre-specified level ($0\!<\!\pi^\star\!<\!1$; commonly: $\pi^\star=33\%$ or $\pi^\star=25\%$). DLTs are usually composite endpoints comprising a range of adverse events according their severity grade.
  
  These exploratory early-phase studies are typically composed by a small sample of healthy volunteers, or patients (for example, in oncology bacause of the potential high toxicity of drugs, or in paediatrics because of the vulnerable population).\citep{chevret06,DaimonHirakawaMatsui} Due to the limited sample size and ethical concerns, simple randomized designs are usually not  applied, but rather response-adaptive sequential designs, including some methods that can potentially find the MTD sooner and limit the number of observed DLTs.\citep{letourneau09,NeuenschwanderEtAl2015} These sequential designs can be divided in two main groups: (1) \emph{algorithm-based} and (2) \emph{model-based} methods. In the first group we can find the classical \emph{\mbox{$3+3$}~design}, which uses prespecified fixed rules and only information of the patients at one dose (the one given to the previous cohort of patients) to allocate the next cohort to a dose level.\citep{ReinerPaolettiOQuigley1999,PaolettiEtAl2015}
  On the other hand, model-based designs, such as the \emph{continual reassessment method (CRM)}\citep{OQuigleyPepeFisher1990} or the \emph{Bayesian logistic regression model (BLRM)},\citep{NeuenschwanderEtAl2008} use an underlying statistical working model to describe the dose-toxicity relationship and a prespecified toxicity rate (the \emph{target toxicity}), to allocate the next cohort of patients.
  While the former are easier to implement, the latter have better operating characteristics, that is, a higher percentage of correct MTD selection and greater chances of promoting effective drugs to subsequent development phases.\citep{onar09,ConawayPetroni2019,WeiPanFernandez2019} As a compromise between these two approaches, \emph{model-assisted} designs have been proposed, which allow to combine the simplicity of rule-based designs with the estimation of model-based designs.\citep{ZhouEtAl2018,YuanLeeHilsenbeck2019}
  Like the model-based approaches, these use a statistical model to develop the design, but fixed dose allocation rules are determined at the pre-planning stage, which makes implementation about as straightforward as for rule-based designs.
  
  In all medical fields, and above all in oncology, identifying the appropriate therapeutical dose range is crucial to avoid exposing patients to an unacceptable toxicity profile or to insufficient efficacy.\citep{bretz05} Recently, Harrison\citep{Harrison2016} reported that phase~II and phase~III study failures were often due to lack of efficacy (52\%) and safety (24\%). They failed owing to inadequate patient selection, study design, biomarkers, and schedules, as well as data analysis, but the most evident resaon was wrong estimation of the dose-response relationship during exploratory clinical trials. Thus, an improper dose selection at the first stages of clinical development may lead either to abandonment of promising drugs, or to the focusing on a wrong dose (too toxic or lacking efficacy). Therefore, using all available data, including other exploratory trials, in order to improve the estimation of the MTD is of utmost importance and the use of meta-analysis methods could be helpful.
  To cope with the specifics of phase~I trials, i.e., small sample sizes and the consideration of several dose levels, only few meta-analytic methods have been proposed so far. Zohar \emph{et~al.}\citep{ZoharEtAl2011} proposed a common-effect (fixed-effect) method based on the retroscpective CRM\@. Ursino \emph{et~al.}\citep{UrsinoEtAl2021} developed a more sophisticated one-stage random-effects approach based on stochastic process priors. 
  Other authors, instead, focused directly on efficacy endpoints, such as Kim \emph{et~al.}\citep{kim17} who developed random-effects meta-analysis methodology to deal with a relaxed exchangeability assumption and rare events. 
  
  Our take was to propose the adaptation and the use of simpler approaches developed for regular meta-analysis, and to check their suitability in realistic phase~I scenarios.
  We considered estimates of the MTD along with their standard errors, which are readily derived from the outcomes of a dose-finding trial based on a simple logistic model. 
  In order to account for the small sample sizes typically encountered in early-phase studies, we suggest the use of a penalization approach correcting for bias in the logistic regression stage.\citep{KosmidisFirth2021,PuhrEtAl2017}
  We then utilize a univariate two-stage meta-analysis model to synthesize the different estimates.
  Implementation is based on the \texttt{logistf} and \texttt{bayesmeta} \textsf{R}~packages.\citep{R:logistf,bayesmeta}
  The results of the proposed meta-analysis methods may be utilized not only for an overall summary estimate (e.g., for consideration by regulatory agencies), but also for the design of a prospective trial,\citep{DerSimonian1996,GoudieEtAl2010} for the selection of the dose panel, for sample size determination,\citep{DeSantis2007} or as a meta-analytic predictive (MAP) prior for Bayesian inference.\citep{SchmidliEtAl2014,WandelNeuenschwanderRoeverFriede2017}
  
  The organisation of this article is as follows; in Section~\ref{sec:examples1}, we motivate the problem by considering two example data sets; in both cases, a literature search on \texttt{clinicaltrials.gov} and PubMed yielded a number of phase~I studies investigating a novel drug at a range of dose levels. In Section~\ref{sec:methods} we introduce the statistical methods for MTD estimation and meta-analysis used in the following. In Section~\ref{sec:simul}, we investigate the proposed methods' performance using Monte Carlo simulations considering the combined estimate as well as shrinkage and prediction. 
  In Section~\ref{sec:examples2}, we analyze the example data, and Section~\ref{sec:conclusions} finally closes with some conclusions.

\section{Illustration}\label{sec:examples1}
  \subsection{Sorafenib example}  
  The first illustrating example concerns Sorafenib (BAY 43-9006), a kinase inhibitor for the treatment of advanced renal cell carcinoma, hepatocellular carcinoma, and radioactive iodine resistant advanced thyroid carcinoma.\citep{EMA-Sorafenib} Thirteen trials with published results, described in 11 manuscripts, were identified in a literature search. The doses used in each trial, the numbers of patients allocated, and the total numbers of DLTs at each dose are summarized in Table~\ref{tab:sorafenibData}.

  \begin{table}[ht]
    \caption{The results of 13 studies on Sorafenib monotherapy. For
      each dose considered in each trial, the numbers of patients
      experiencing DLT events, and the total numbers of exposed
      patients are given.}\label{tab:sorafenibData}
    \begin{center}
    \begin{tabular}{lllccccccc}
      \toprule
      &&& \multicolumn{7}{c}{dose (mg)}  \\
      \cmidrule(lr){4-10}
      publication                  & year & country\hspace{3ex} & 100 &  200 & 300 &  400 &  600 & 800 & 1000 \\
      \midrule
      Awada \emph{et~al.} \citep{AwadaEtAl2005}         & 2005 & Belgium & 0/4 &  0/3 & 1/5 & 1/10 & 7/12 & 1/3 & \\
      Clark \emph{et~al.} \citep{ClarkEtAl2005}         & 2005 & USA     & 0/3 &  0/3 &     &  1/4 &  1/6 & 3/3 & \\
      Moore \emph{et~al.} \citep{MooreEtAl2005}         & 2005 & Canada  & 0/3 &  1/6 &     &  0/8 &  3/7 &     & \\
      Strumberg \emph{et~al.} \citep{StrumbergEtAl2005} & 2005 & Germany & 1/5 &  1/6 &     & 0/15 & 4/14 & 2/7 & \\
      Furuse \emph{et~al.} \citep{FuruseEtAl2008}       & 2008 & Japan   &     & 0/12 &     & 1/14 &      &     & \\[1ex]
      Minami \emph{et~al.} \citep{MinamiEtAl2008}       & 2008 & Japan   & 0/3 & 1/12 &     &  0/6 &  1/6 &     & \\
      Miller \emph{et~al.} \citep{MillerEtAl2009}       & 2009 & USA     &     & 8/34 &     & 6/20 &      &     & \\
      Crump \emph{et~al.} (A) \citep{CrumpEtAl2010}     & 2010 & Canada  & 0/4 &  1/6 & 0/6 &  1/6 &      &     & \\
      Crump \emph{et~al.} (B) \citep{CrumpEtAl2010}     & 2010 & Canada  & 0/3 &  1/6 &     &  0/3 &  2/6 &     & \\
      Borthakur \emph{et~al.} (A) \citep{BorthakurEtAl2011} & 2011 & USA     &     &  0/3 &     & 0/15 &  2/8 &     & \\[1ex]
      Borthakur \emph{et~al.} (B) \citep{BorthakurEtAl2011} & 2011 & USA     &     &  0/3 &     &  1/7 &  2/6 &     & \\
      Nabors \emph{et~al.} \citep{NaborsEtAl2011}       & 2011 & USA     &     &  0/3 &     &  1/6 &  0/3 & 1/5 & 3/3 \\
      Chen \emph{et~al.} \citep{ChenEtAl2014}           & 2014 & USA     &     &  0/3 &     & 1/16 &      &     & \\
    \bottomrule
    \end{tabular}
    \end{center}
  \end{table}

  Seven doses (100, 200, 300, 400,  600, 800 and 1000 mg) were investigated across all studies. DLT definitions and Sorafenib administration schedules (about a 28-day cycle) were comparable across studies.
  The total sample sizes vary from~16 up to~54 patients, the maximum was achieved in one of the studies where Sorafenib was tested in 9~different patient subgroups defined by organ dysfunctions.\citep{MillerEtAl2009} As expected for dose-escalation trials, observed DLT frequencies generally tend to increase with increasing dose, with zero or a low event rates at the lower doses placed at the beginning of the dose panel.

  The majority of studies was performed in European or North American populations, while only two studies were done in Japanese patients.\citep{FuruseEtAl2008,MinamiEtAl2008} 
  Indeed, the Japanese Pharmaceuticals and Medical Devices Agency (PMDA) recommends a treatment re-evaluation in case insufficient data are available from a Japanese population.\citep{pmda}
  As it is unclear to what extent evidence from the differing (Western and Japanese) populations is directly transferrable to one another,\citep{ICH1998,EMA2013} we can consider a cautious \emph{dynamical borrowing} approach in order to bridge between the two study subsets rather than pooling the data ``na\"{i}vely''.\citep{RoeverFriede2020}

  \subsection{Irinotecan / S-1 example}
  The second illustrating example concerns a combination therapy of Irinotecan (a topoisomerase~1 inhibitor) and \mbox{S-1} (a combination of three pharmacological compounds, namely, tegafur, gimeracil, and oteracil potassium) that was tested in advanced colorectal and gastric cancer in a Japanese population.\citep{EMA-Irinotecan,EMA-S1}
  Data extracted from 12~studies are shown in Table~\ref{tab:irinotecanData}.

  \begin{table}[ht]
    \caption{The results of 12~Japanese studies on combination therapy of Irinotecan and \mbox{S-1}. For each dose considered in each trial, the numbers of patients experiencing DLT~events, and the total numbers of exposed patients are given.}
    \label{tab:irinotecanData}
    \begin{center}
    \begin{tabular}{llcccccccccc} 
      \toprule
      && \multicolumn{10}{c}{dose (mg/m$^2$)}  \\
      \cmidrule(lr){3-12}
      publication              & year\hspace{2ex} & 40  &  50 &   60 &  70 &    80 &  90 & 100 & 120 & 125 & 150 \\
      \midrule
      Yamada \emph{et~al.} \citep{YamadaEtAl2003}   & 2003 &     &     &      &     &       &     & 0/3 &     & 0/3 & 1/6 \\
      Takiuchi \emph{et~al.} \citep{TakiuchiEtAl2005} & 2005 & 1/6 &     &  0/3 &     &   0/4 &     & 3/6 &     &     & \\
      Inokuchi \emph{et~al.} \citep{InokuchiEtAl2006} & 2006 &     &     &      & 0/3 & 10/42 & 0/3 & 2/3 &     &     & \\
      Nakafusa \emph{et~al.} \citep{NakafusaEtAl2008} & 2008 &     &     & 7/39 &     &   2/3 &     &     &     &     & \\
      Ishimoto \emph{et~al.} \citep{IshimotoEtAl2009} & 2009 &     & 0/3 &  0/3 & 0/3 &   2/4 &     &     &     &     & \\[1ex]
      Ogata \emph{et~al.} \citep{OgataEtAl2009}    & 2009 & 0/3 & 0/3 &  3/4 &     &       &     &     &     &     & \\
      Shiozawa \emph{et~al.} \citep{ShiozawaEtAl2009} & 2009 &     &     &      &     &   1/6 &     & 2/6 & 2/6 &     & 2/3 \\
      Yoshioka \emph{et~al.} \citep{YoshiokaEtAl2009} & 2009 &     &     &      &     &       &     & 0/3 &     & 1/6 & 0/3 \\
      Komatsu \emph{et~al.} \citep{KomatsuEtAl2010}  & 2010 &     &     &      &     &       &     & 1/9 &     & 1/9 & 0/3 \\
      Kusaba \emph{et~al.} \citep{KusabaEtAl2010}   & 2010 &     &     &      &     &   0/6 &     & 2/3 &     &     & \\[1ex]
      Yoda \emph{et~al.} \citep{YodaEtAl2011}     & 2011 &     &     &  0/3 &     &   3/6 &     &     &     &     & \\
      Goya \emph{et~al.} \citep{GoyaEtAl2012}     & 2012 &     &     &      & 0/3 &   0/3 & 3/5 &     &     &     & \\
      \bottomrule
    \end{tabular}
    \end{center}
  \end{table}

  A total of 10~doses, ranging from 40 up to 150 mg/m$^2$, were evaluated among all trials. Two or more infusions of Irinotecan were planned in all trials, except for Yamada \emph{et~al.}\citep{YamadaEtAl2003} and Yoshioka \emph{et~al.}\citep{YoshiokaEtAl2009} with only one infusion at the first day cycle. The sample sizes range from~6, in agreement with the well-known \mbox{3+3}~design, to~51, since, in Inokuchi \emph{et~al.},\citep{InokuchiEtAl2006} DLTs were still recorded at the phase~II stage of the study.

\section{Methods}\label{sec:methods}
  \subsection{Dose-finding experiments: assumptions}
    Within a dose-finding experiment, we have a discrete set of covariable levels~$x_j$, usually denoted as \emph{doses} or \emph{exposures}, or transformations thereof, which are indexed in increasing order. The response is an event count~$r_j$ among a total number~$n_j$ of patients that have been exposed to the $j$th dose.

    We assume the DLT count~$r_j$ among the total of $n_j$~patients exposed at level~$x_j$ to follow a binomial distribution,
    \begin{eqnarray}\label{eqn:binomial}
      r_j & \sim & \binomdistn(p_j ,\, n_j)
      \mbox{.}
    \end{eqnarray}
    MTD estimation then aims at determining the dose $x^\star$ for which the DLT probability reaches a certain 
    threshold $\pi^\star$
    (or sometimes also the largest dose with DLT probability $p_j \leq \pi^\star$). 
    Depending on the approach taken, the dose level~$x^\star$ may be assumed to be among the set of experimentally considered dose levels ($x_1,\ldots,x_J$) or also in between or beyond the investigated doses.

  \subsection{Logistic regression model}
    We apply logistic regression models with a logit link for each study. The DLT~counts are modelled via a binomial distribution as in \eqref{eqn:binomial}. The logistic regression model provides a joint parametrization of the DLT probabilities~$p_j$ as a parametric function of the dose covariable~$x_j$:
    \begin{equation}\label{eqn:logitModel}
      \logit(p_j) \; = \; \beta_0  + \beta_1 x_j
    \end{equation}
    where the logit link function and its inverse are defined as
    \begin{equation} \textstyle
      \logit(x) \; = \; \log\bigl(\frac{x}{1-x}\bigr)
      \qquad \mbox{and} \qquad
      \logit^{-1}(y) \; = \; \frac{\exp(y)}{1+\exp(y)}
      \mbox{.}
    \end{equation}
    Note that the \emph{logits} of the probabilities corresponds to the \emph{logarithmic odds}, where the \emph{odds} correponding to a probability~$p\in[0,1]$ are given by $\frac{p}{1-p} \in [0,\infty]$. The linearity assumption constrains the DLT probabilities and reduces dimensionality of the problem: only 2~unknown parameters, the intercept~$\beta_0$ and slope~$\beta_1$ now determine the dose-response relationship. In addition, it facilitates interpolation or extrapolation beyond the discrete set of covariable ($x_j$)~values. Introduction of a parametric model also means that monotonic transformations of the dose levels~($x_i$) may affect the model fit (see also Figure~\ref{fig:ExampleScenarios}).

    Within the logistic model, the MTD directly results from the regression parameters ($\beta_0$ and $\beta_1$).  Inverting equation~\eqref{eqn:logitModel} to derive the dose level~$x^\star$ for which the pre-specified toxicity of~$\pi^\star$ is attained (i.e., so that $\beta_0+\beta_1x^\star = \logit(\pi^\star)$) leads to
    \begin{equation}\label{eqn:logisticMTD}
      x^\star \;=\;  \frac{\logit(\pi^\star)-\beta_0}{\beta_1}
    \end{equation}
    as the corresponding MTD\@.

    The logistic regression model as a very common procedure is a simple and pragmatic model choice here; a number of other models or parametrisations are also commonly used in dose-response modeling.\citep{BretzEtAl2005,PinheiroEtAl2014} The logistic model also includes the so-called \emph{Emax~model} as a special case (see Appendix~\ref{sec:emaxAppendix} for details).

  \subsection{Parameter estimation in the logistic model}
    Simple maximum-likelihood (ML) estimation within the logistic regression model is sometimes problematic, in particular in case of smaller sample sizes. ML~estimators are affected by certain biases, and \emph{separation} issues may arise.\citep{AlbertAnderson1984} Separation refers to ``pathological'' data constellations where the predictor allows to perfectly split the data into cases and non-cases, which in the estimation stage leads to problems (a likelihood that doesn't attain a definite maximum at finite parameter values, which in practice often leads to numerical problems, resulting in large estimates as well as associated standard errors).

    The use of the \emph{Firth correction} has been suggested in order to reduce bias and to ensure finite parameter estimates.\citep{Firth1993,KosmidisFirth2021} Heinze and colleagues\citep{HeinzeSchemper2002,HeinzePuhr2010} showed that the use of the Firth correction avoids separation issues and may improve properties also in small samples. In the present context, our focus is not only on contrast estimates (odds ratios) but also hinges on the accuracy absolute (DLT) event probability estimates. Puhr \emph{et~al.}\citep{PuhrEtAl2017} introduced \emph{Firth's logistic regression with added covariate (FLAC)} as a method to further reduce bias in event probability estimates; the FLAC method has been recommended due to improved performance as well as its invariance properties.\citep{PuhrEtAl2017}

    The logistic model falls within the class of \emph{generalized linear models (GLMs)}, and as such, implementations are readily available. Within \textsf{R}, one may utilize the ``\texttt{glm()}'' function for simple logistic regression, while the ``\texttt{MASS}'' library also provides the functionality to retrieve estimates and standard errors for the MTD\@.\citep{VenablesRipleyCh7} Firth logistic regression and FLAC are implemented in the ``\texttt{logistf}'' \textsf{R}~package.\citep{R:logistf} Point estimates and their associated variance-covariance may again be utilized to derive corresponding MTD estimates and standard errors.

  \subsection{Deriving (MTD) dose estimates}
    \subsubsection{General remarks}
      Note that the MTD~$x^\star$ does not necessarily fall within the range of investigated doses ($x_1,\ldots,x_j$); the resulting estimate hence is usually an interpolation, or even an extrapolation. The actual task is to use the inverted equation~(\ref{eqn:logisticMTD}) based on \emph{estimates} of the coefficients~$\beta_0$ and~$\beta_1$ in order to derive an estimate of the MTD~$x^\star$.  A common approach is to assume a joint normal distribution for the ML estimates~$\hat{\beta}_0$ and~$\hat{\beta}_1$. The distribution of the estimated MTD then results as a ratio of (correlated) normal variates (see~(\ref{eqn:logisticMTD})).
      Such a ratio may often again be reasonably approximated by a normal distribution, however, in general the distribution may also be bimodal and heavy-tailed.\citep{PhamGiaEtAl2006} In particular, its first two moments (mean and variance) in general do not exist.

      In the following, we will investigate a simple error propagation approach based on the \emph{delta method}. Note that the eventual objective of this exercise is to provide estimates and standard errors to be passed on to the meta-analysis procedure at a subsequent stage. Estimates and standard errors are then utilized to specify the (approximately) normal likelihood (see Section~\ref{sec:meta} below). The aim hence is to provide an accurate (or reasonably conservative) normal approximation to the uncertainty in the MTD\@.\citep{Bretthorst1999}

    \subsubsection{The delta method}
      Consideration of standard errors (the variance-covariance matrix) of the regression coefficients~$\beta_0$ and~$\beta_1$ via the \emph{delta method} yields a corresponding standard error for the MTD estimate~$x^\star$.\citep{Cox2005} This procedure is described in detail, including \textsf{R}~code, by Venables and Ripley.\citep{VenablesRipleyCh7} It should be noted, however, that this only works well in case of reasonably small standard errors; in particular, the denominator in equation~(\ref{eqn:logisticMTD}), the regression slope, needs to be somewhat bounded away from zero (i.e., it needs to have a small coefficient of variation).\citep{PhamGiaEtAl2006}

  \subsection{Two-stage approach to meta-analysis}\label{sec:meta}
    \subsubsection{The normal-normal hierarchical model (NNHM)}
      With MTD estimates that are provided along with their standard errors, the meta-analysis problem falls into the generic class of problems that may be addressed using the normal-normal hierarchical model (NNHM)\@.\citep{HedgesOlkin,HartungKnappSinha,BorensteinEtAl,BorensteinEtAl2010,Viechtbauer2010,Roever2020}
      For each study~$i$ ($i=1,\ldots,k$), there is an underlying (unknown) true MTD~value~$\theta_i$.
      Analysis of the experimental data yields an MTD~estimate~$y_i$ that has a standard error~$s_i$ associated. 
      Utilizing a simple normal model, we assume that
      \begin{equation}\label{eqn:NNHM1}
        y_i|\theta_i,s_i \; \sim \; \normaldistn(\theta_i, s_i^2) \mbox{,}
      \end{equation}
      where the study-specific true values~$\theta_i$ (the MTDs) are not necessarily identical across the different studies; these also have a certain amount of variability associated, which is implemented using another variance component via
      \begin{equation}\label{eqn:NNHM2}
        \theta_i|\mu,\tau \; \sim \; \normaldistn(\mu,\tau^2) \mbox{.}
      \end{equation}
      The heterogeneity parameter~$\tau$ denotes the between-study variability, and $\mu$ is the overall mean. The model may also be expressed in its marginal form as
      \begin{eqnarray}\label{eqn:NNHM3}
        y_i|\mu,\tau,s_i & \sim & \normaldistn(\mu,\,s_i^2+\tau^2) \mbox{.}
      \end{eqnarray}

    \subsubsection{Estimation within the NNHM}\label{sec:nnhm}
      In the Bayesian framework, the estimates~$y_i$ and their standard errors~$s_i$ are utilized to define the (approximately) normal data likelihood; inference needs to consider the unknowns~$\mu$ and~$\tau$, and primary interest is usually in the overall mean~$\mu$.
      For the design and analysis of a future dose-finding study, prediction (of a ``new'' MTD~$\theta_{k+1}$) or shrinkage estimation (of one of the MTDs~$\theta_i$) may also be relevant.\citep{SchmidliEtAl2014,WandelNeuenschwanderRoeverFriede2017}
      Prior distributions need to be specified for the overall mean~($\mu$) and the heterogeneity~($\tau$), experssing any a-priori information that may be available on these parameters.
      In the following, we will utilize uniform effect and heterogeneity priors, due to the mostly large sample sizes~(numbers of studies~$k$) in the examples. In general, or in particular when faced with only a small number of studies, the use of (weakly) informative priors may also be appropriate.\citep{Roever2020,RoeverEtAl2021}

\section{Simulations}\label{sec:simul}
  \subsection{Aims}
    The aim is to firstly investigate the performance of the (1st-stage) regression methods, and then also of the (2nd-stage) meta-analysis.  For the logistic regression models, we will check bias, CI coverage probabilities and CI widths; for the meta-analyses, we focus on CI coverages and widths (for overall mean, prediction, shrinkage estimates). To this end, we will utilize Monte Carlo expriments using simulated data from a range of relevant scenarios. 

    Even when assuming that the logistic regression analyses yield relatively unbiased estimates of the regression coefficients, the eventual consideration of derived MTDs (a \emph{nonlinear} transformation, see equation~(\ref{eqn:logisticMTD})) means that we again should expect some bias. Considered as a function of the slope~$\beta_1$, the transformation in~(\ref{eqn:logisticMTD}) is a \emph{convex} function, so if we assume that the MTD uncertainty was dominated by uncertainty in the slope, then (following Jensen's inequality) we may expect a negative bias here. Considered as a function of both intercept and slope, however, the MTD is neither convex nor concave.
 
    In addition, it also remains to be seen how well a normal approximation works in a setup where moments may actually not be finite.\citep{PhamGiaEtAl2006}

  \subsection{Simulation scenarios}
\begin{figure}
      \centering
      \includegraphics[width=0.95\textwidth]{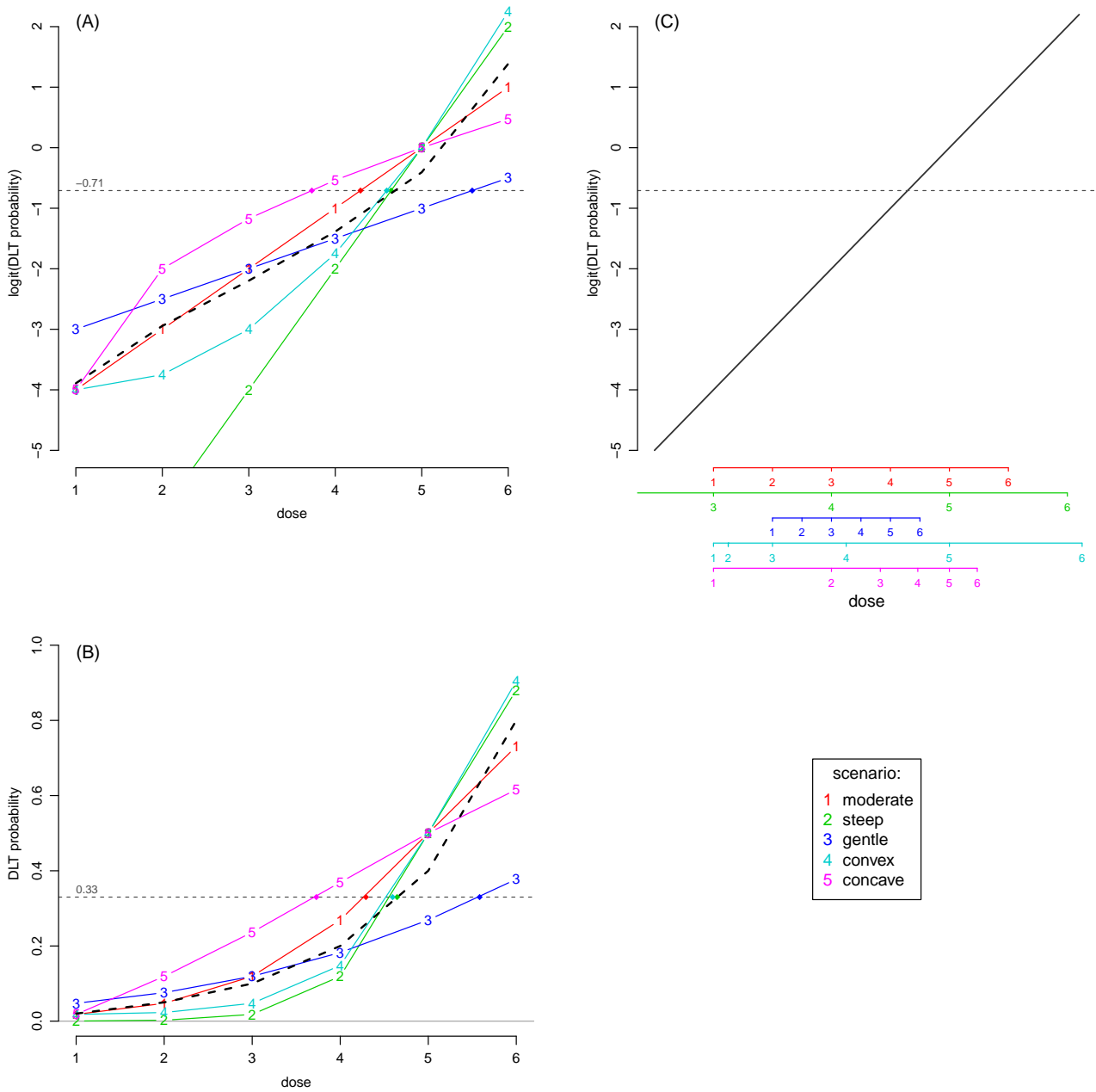}
      \caption{The five simulation scenarios (dose-response curves) on the probability (A) and logit scales (B). The dashed line indicates the ``skeleton'' (a prior guess of the curve, which needs to be specified for CRM and BLRM methods). The corresponding MTDs, the doses at which a probabilitiy of~$0.33$ is attained (where $\logit(0.33)=-0.71$), are marked by a dot.
      See also Table~\ref{tab:ExampleScenarios} for the actual numbers.
      Panel~(C) illustrates that the five scenarios may also be motivated via a common underlying dose-response curve, but differing sets of dose levels.}
      \label{fig:ExampleScenarios}
    \end{figure}

    The simulation scenarios considered are intended to be realistic in terms of the data that are commonly encountered in dose-escalation studies (as, e.g., in the examples from Section~\ref{sec:examples1}; see also Appendix~\ref{sec:ExampleDataAppendix}). 
    We will use a mixture of data generated by employing the common ``\mbox{3+3}'',\citep{ReinerPaolettiOQuigley1999,PaolettiEtAl2015} ``CRM'',\citep{OQuigleyPepeFisher1990,GarrettMayer2006} and ``BLRM''\citep{NeuenschwanderEtAl2008,SweetingManderSabin2013,NeuenschwanderEtAl2015} study designs (with equal probabilities).
    While the algorithmic \mbox{3+3}~design used to be the most common in the past, alternative designs have recently gained popularity.\citep{RiviereEtAl2015,VanBrummelenEtAl2016} 
    The different designs will then result in differing kinds of data sets, in terms of sample sizes, event rates, investigated dose ranges, etc.\citep{LinShih2001,KangAhn2002}
    Implementation details regarding the three designs are provided in Appendix~\ref{sec:DataGenerationAppendix}.
    \begin{description}
      \item[Toxicity profiles:] we will use 5 dose-response curves that are shown in Figure~\ref{fig:ExampleScenarios} and Table~\ref{tab:ExampleScenarios}. All of these are based on~6 dose levels ($x_i\in\{1,2,\ldots,6\}$), and they are labeled as ``moderate'', ``steep'', ``gentle'', ``convex'' or``concave''. 
      The first three are linear, while the latter two deviate from the logit model assumed in the analyses.
            Defining the targeted DLT probability as $\pi^\star=0.33$, the MTD is attained within the dose range (between~1 and~6) in all 5~cases; the MTDs are also listed in Table~\ref{tab:ExampleScenarios}.
      \item[Design criteria:] the differing dose-escalation designs require the specification of certain free parameters. The \mbox{3+3} design is initiated at the lowest dose, a cohort size of~3 is used, there is no skipping of doses, and no dose escalation following a DLT~event. CRM and BLRM designs utilize the ``skeleton'' (a prior guess of the response curve) shown in Figure~\ref{fig:ExampleScenarios} and Table~\ref{tab:ExampleScenarios}. For CRM and BLRM, the maximum sample size was chosen randomly as a multiple of~3 between~15 and~30.
      \item[Heterogeneity:] Heterogeneity in MTDs will be implemented via variability in the \emph{intercept} term of the dose-response curve. In Figure~\ref{fig:ExampleScenarios}, one can see that shifting the curves in the vertical direction will also increase or decrease the associated MTD\@. For a logistic dose-response (as in equation~\ref{eqn:logitModel}), adding a random offset with variance~$(\tau\beta_1)^2$ will imply a heterogeneity variance of~$\tau^2$ in the associated MTD\@. For the non-linear scenarios, the relationship is not as simple, but instead of the slope~$\beta_1$, the slope \emph{at the MTD} (as in Table~\ref{tab:ExampleScenarios}) is simply used instead as an approximation. Heterogeneity levels of $\tau\in\{0.0, 0.5, 1.0\}$ will be considered.
      \item[Numbers of studies:] At the meta-analysis stage, totals of $k\in\{5,10,20\}$ studies will be considered.
    \end{description}
    It should be noted that the five different dose-response curves may also be related to one another by considering a ``common'' curve with differing dose placement; see also the illustration in Figure~\ref{fig:ExampleScenarios}. 
    Panel~(C) shows how the five scenarios may result from considering different sets of investigated doses. A denser or wider spacing affects the resulting slope, and constancy of distances between neighbouring doses affects overall linearity. 
    Differences between scenarios do not only relate to the underlying dose-response mechanism, but also to choices made at the experimental design stage. For example, the third (``gentle'') scenario may be viewed as a scenario in which only a very narrow dose range is investigated --- which eventually (and maybe as expected) makes it harder to identify the exact MTD, while at the same time problems appear amplified since uncertainty in the MTD in terms of a certain absolute dose range then quickly translates into uncertainty spanning several dose levels.
    This highlights the importance of the careful choice of the dose (or \emph{exposure}) scale and of possible transformations (e.g. when considering log-transformed doses).
    It also  emphasizes the difficulties arising in the comparison of results (in particular: standard errors and CI widths quoted in units of \emph{doses}) across the different simulation scenarios. Seemingly ``steep'' and ``gentle'' shapes may simply arise from from considering wider or narrower dose ranges.

     \begin{table}[ht]
      \caption{The DLT probabilities in the five example scenarios, and the associated true MTDs. See also Figure~\ref{fig:ExampleScenarios}.}\label{tab:ExampleScenarios}
      \begin{center}
        \begin{tabular}{llcccccc}
          \toprule
          && \multicolumn{5}{c}{scenario} &  \\
          \cmidrule(lr){3-7}
          & & \textit{moderate} & \textit{steep} & \textit{gentle} &  \textit{convex} & \textit{concave} &  skeleton \\
          \midrule
          dose & 1 & 0.018 & 0.00034 & 0.047 & 0.018 & 0.018 & 0.02 \\
               & 2 & 0.047 & 0.0025\phantom{0} & 0.076 & 0.023 & 0.119 & 0.05 \\
               & 3 & 0.119 & 0.018\phantom{00} & 0.119 & 0.047 & 0.237 & 0.10 \\
               & 4 & 0.269 & 0.119\phantom{00} & 0.182 & 0.148 & 0.369 & 0.20 \\
               & 5 & 0.500 & 0.500\phantom{00} & 0.269 & 0.500 & 0.500 & 0.40 \\
               & 6 & 0.731 & 0.881\phantom{00} & 0.378 & 0.905 & 0.616 & 0.80 \\[1ex]
          MTD  &   & 4.29\phantom{0} & 4.65\phantom{000} & 5.58\phantom{0} & 4.60\phantom{0} & 3.73\phantom{0} &      \\
          \bottomrule
        \end{tabular}
      \end{center}
    \end{table}

  \subsection{Simulation scenario properties}
     We briefly investigated the kinds of data that are returned when implementing the above model assumptions. Table~\ref{tab:ExampleDataProperties} characterizes the data in terms of the mean numbers of doses used, the numbers of patients recruited and the numbers of DLT events under the three designs and overall (on average across designs).

    \begin{table}[ht]
      \caption{Mean numbers of doses / patients / events resulting from the different data scenarios and designs. Note that for CRM and BLRM, the maximum sample size is also random here (between~15 and~30).}\label{tab:ExampleDataProperties}
      \begin{center}
        \begin{tabular}{lccccc}
          \toprule
          & \multicolumn{5}{c}{scenario} \\
          \cmidrule(lr){2-6}
          design & \textit{moderate} & \textit{steep} & \textit{gentle} &  \textit{convex} & \textit{concave} \\
          \midrule
          3+3  & 4.3 / 16.1 / 2.8 & 5.0 / 17.1 / 2.7 & 4.6 / 17.4 / 2.7 & 4.9 / 17.2 / 2.8 & 3.7 / 13.9 / 2.7 \\
          CRM  & 4.5 / 22.3 / 4.6 & 5.0 / 22.5 / 4.8 & 4.6 / 22.6 / 3.5 & 4.8 / 22.5 / 4.7 & 4.0 / 22.5 / 5.3 \\
          BLRM & 4.5 / 22.6 / 4.8 & 5.0 / 22.6 / 4.0 & 4.9 / 22.8 / 3.8 & 4.9 / 22.5 / 4.2 & 4.0 / 22.4 / 5.4 \\[1ex]
          overall & 4.5 / 20.3 / 4.1 & 5.0 / 20.7 / 3.9 & 4.7 / 20.9 / 3.3 & 4.9 / 20.7 / 3.9 & 3.9 / 19.6 / 4.5\\
          \bottomrule
        \end{tabular}
      \end{center}
    \end{table}

    Most notably, CRM and BLRM yield very similar data, while 3+3 differs slightly from the other two. The 3+3 design generally yields a smaller total sample size (numbers of patients \emph{and} events). In the following, we will not distinguish between data generating models, but will consider a mixture of the three. The data characteristics are also roughly similar to those encountered in the Sorafenib and Irinotecan examples above (see also Tables~\ref{tab:sorafenibData}, \ref{tab:irinotecanData}, \ref{tab:sorafenibEstimates} and~\ref{tab:irinotecanEstimates}).

  \subsection{Results}\label{sec:Results}
    \subsubsection{Comparing standard and penalized logistic regression methods}
\begin{figure}
        \centering
        \includegraphics[width=0.99\textwidth]{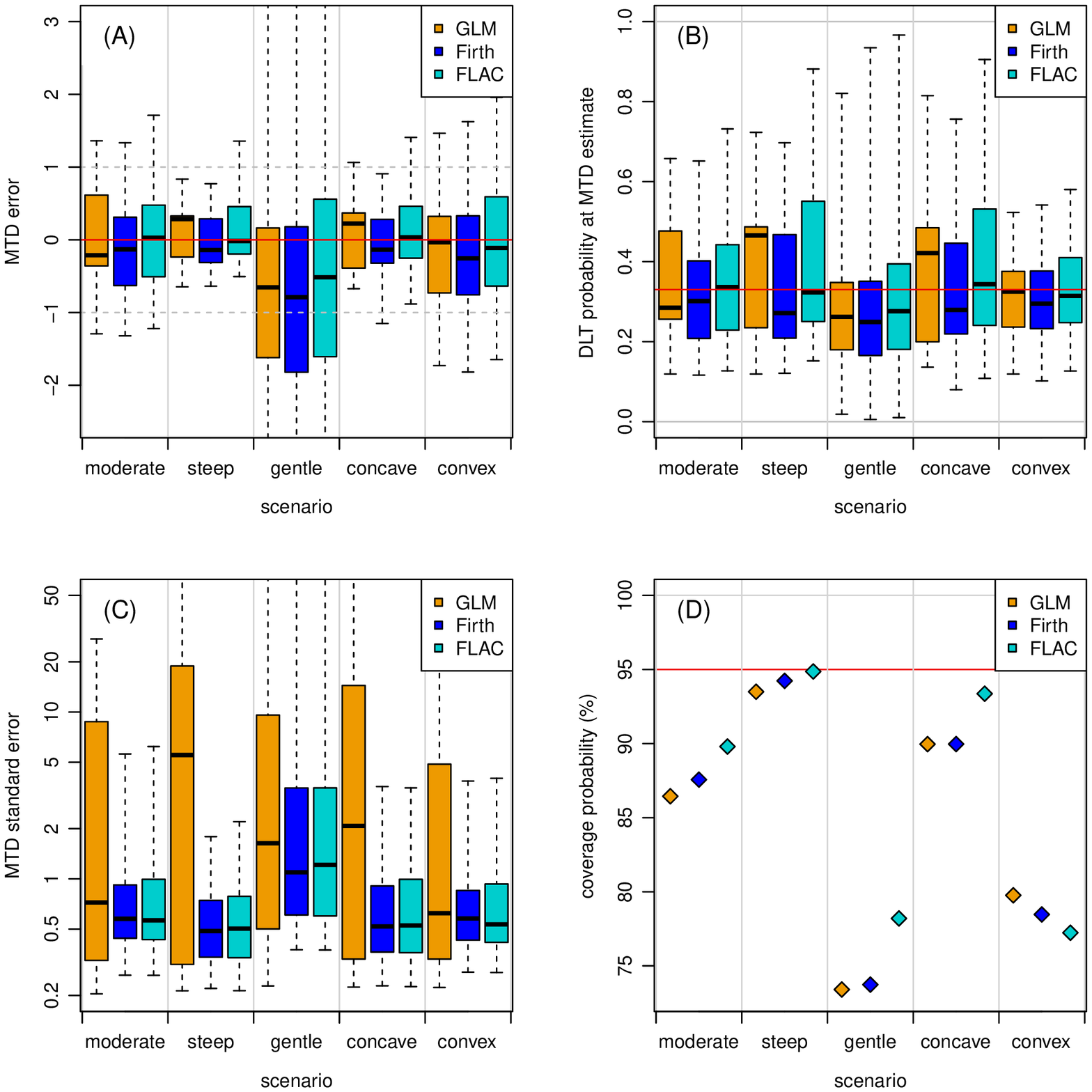}
        \caption{Performance of the first-stage logistic regression methods. The top left panel (A) shows the errors in MTD estimates in terms of the offset in \emph{doses}. The top right panel (B) illustrates the error in terms of the \emph{DLT probability} at the estimated MTD; the horizontal reference line indicates the target of 33\%. Panel~(C) shows the standard errors associated with the differing estimates, and panel~(D) shows the coverage probabilities of derived 95\%~confidence intervals. Boxplots indicate the three quartiles and the central 90\% range.}
        \label{fig:mtd-first-stage}
      \end{figure}

      Figure~\ref{fig:mtd-first-stage} illustrates the estimation performance of the three investigated methods (``plain'' logistic regression within the GLM framework, Firth correction, and FLAC). 
      Panel~(A) shows the error in the estimated MTD; within each scenario, all three methods perform comparably, while FLAC appears to be overall least biased. Overall there appears to be a tendency to underestimate the MTD, i.e., the errors are on the conservative side.
      In particular the third scenario, characterized by a very gentle slope, leads to relatively large offsets (in terms of dose steps) and a negative bias. This may in fact be expected, as this is the scenario in which the slope estimate should tend to be particularly uncertain.
    
      Panel~(B) again illustrates the offset in the MTD estimate, this time in terms of the DLT probability at the estimated MTD\@. On this scale, differences between scenarios do not appear quite as dramatic. The probabilities are centered at their aimed for value of $\pi^\star=0.33$ (marked by a red line), and depending on the scenario, they exhibit more or less variability. The FLAC estimate again appears to be closest to the target on average.
    
      Panel~(C) shows the uncertainties (estimated standard errors) associated with the DLT estimates. The Firth and FLAC estimates behave similarly, while the simple GLM estimate yields more extreme (both very small or very large) errors. Note that the standard error is in units of doses here, i.e., a standard error~$>1$ means that the associated CI will span across several doses.
    
      Panel~(D) eventually shows the 95\%~CIs' coverage probabilities for the different combinations; in general we see some undercoverage, which is not actually satisfactory, while in most scenarios (in particular in those where the linearity assumption is met), the FLAC method again tends to perform best.

      Overall, the regularized (Firth or FLAC) estimates clearly outperform the ``plain'' logistic regression for MTD estimation based on relatively small sample sizes. 
      Also, while ``plain'' regression would fail in about 4\% of cases, the other two always yield finite parameter estimates.
      Firth and FLAC estimates seem to behave more similarly, with a slight advantage apparent for the FLAC method. In the following, we will hence focus on the FLAC method for deriving MTD estimates (and associated standard errors) in the first stage of the analysis.

    \subsubsection{Investigating MA performance}
      After checking the performance of MTD estimators for the first stage of the analysis, we will investigate the performance of meta-analyses based on estimated MTDs. In the meta-analysis setup, we will vary the five dose-response scenarios, as well as the amount of heterogeneity and the number of studies considered. As in the previous section, we will be considering the aspects of estimation error (in terms of doses and in terms of DLT probability), as well as CI width and coverage probability.
    
\begin{figure}
        \centering
        \includegraphics[width=0.85\textwidth]{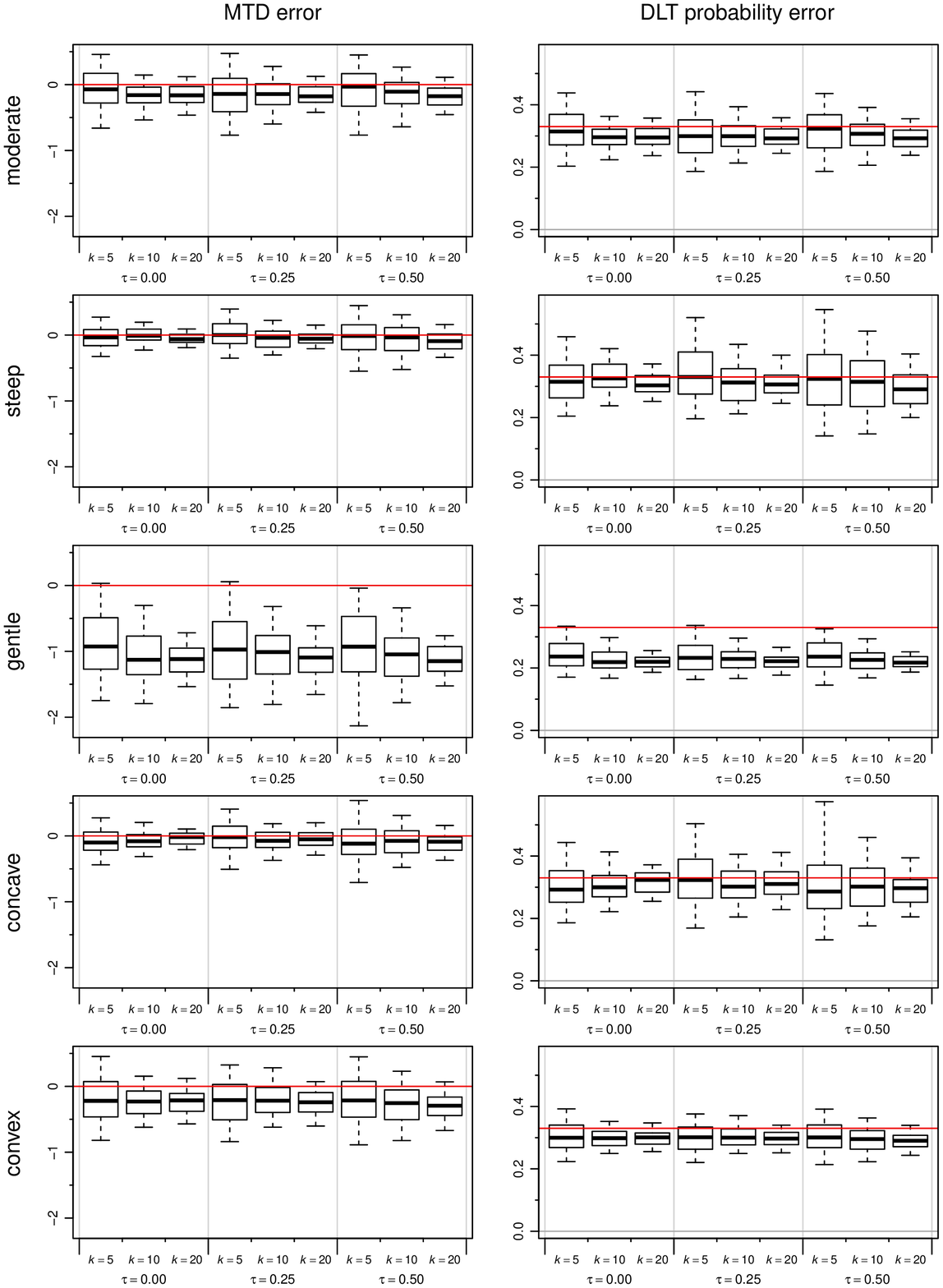}
        \caption{Performance of meta-analysis (based on FLAC estimates) in different scenarios, and for different numbers of studies and amounts of heterogeneity. The left column shows the offset in estimated MTD in terms of \emph{doses}, and the right column shows the DLT probabilities at the estimated MTDs. Rows correspond to the different dose-response scenarios. Boxplots indicate the three quartiles and the central 90\% range.}
        \label{fig:mtd-errors}
      \end{figure}

      Figure~\ref{fig:mtd-errors} illustrates the errors in MTD estimation, both on the dose scale as well as the DLT probability scale. 
      It becomes evident that negative biases already apparent in the previous simulations propagate through to the meta-analysis results. Especially in the ``gentle'' scenario, the MTD gets underestimated, and an increasing number of studies~($k$) reinforces the biased estimate. In this scenario, the MTD estimate is about one dose off, corresponding to a DLT probability of roughly 20--25\% instead of the aimed for~$\pi^\star$ of 33\% (see also Table~\ref{tab:ExampleScenarios}). For all the other scenarios, however, and in particular those where the linearity assumption is met, the estimates appear to behave reasonably, irrespective of the amount of heterogeneity~($\tau$) or the number of studies~($k$).

\begin{figure}[h!tb]
        \centering
        \includegraphics[width=0.85\textwidth]{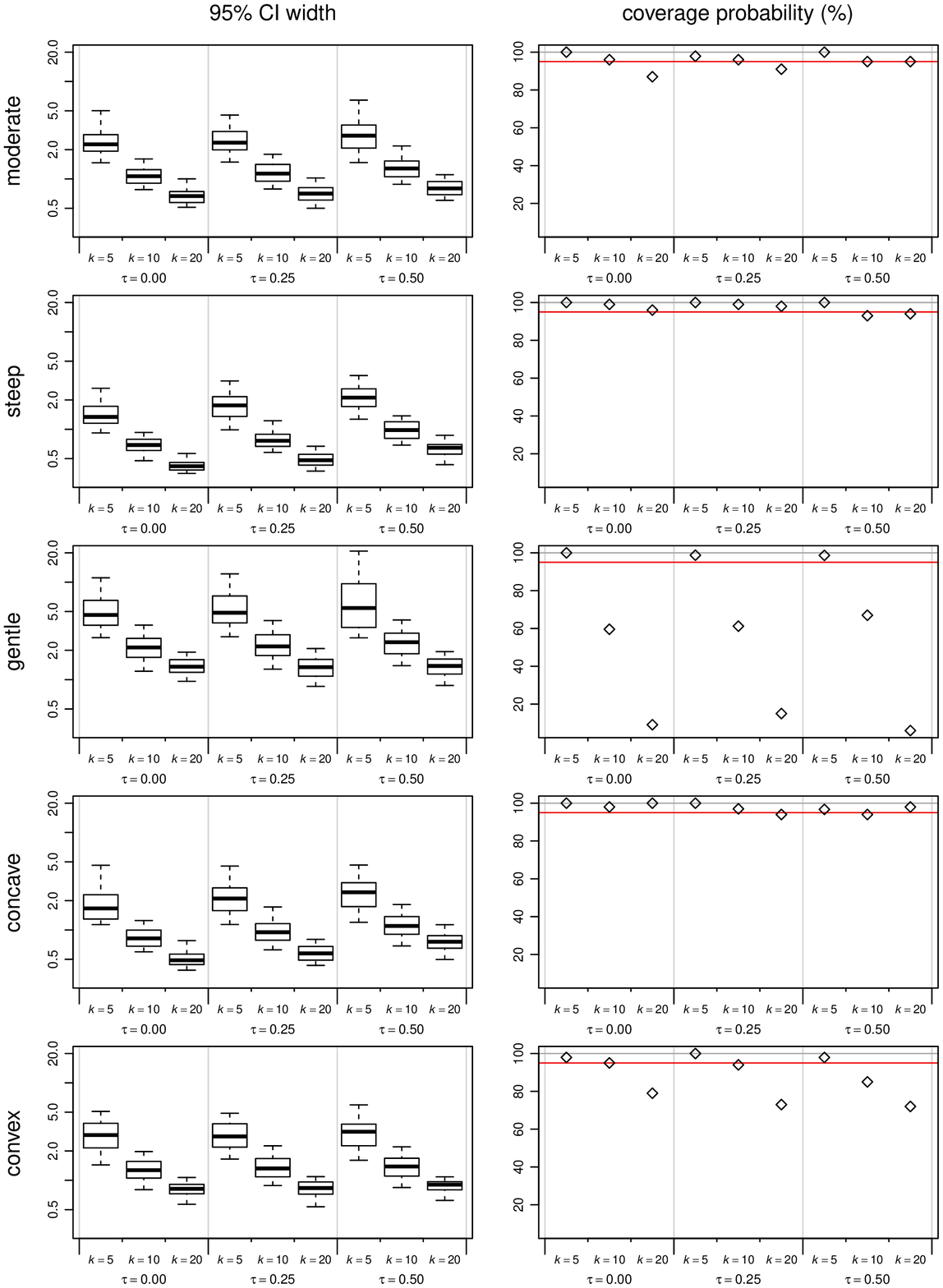}
        \caption{Widths and coverage probabilities of credible intervals for the (mean) MTD based on the meta-analyses. Boxplots indicate the three quartiles and the central 90\% range.}
        \label{fig:ci-mean}
      \end{figure}

      Figure~\ref{fig:ci-mean} illustrates the widths and coverage probabilities of credible intervals for the MTD\@. 
      The length of CIs behaves as expected; an increasing number~($k$) of studies considered leads to shorter intervals, while increasing the amount of heterogeneity~($\tau$) makes intervals slightly wider.
      Coverage probabilities of 95\% intervals however are substantially off in some cases, which makes sense given the previously considered simulation results; when estimation bias is substantial and a large number of studies is included, so that bias is of the order of the CI width, then the coverage deteriorates. In the majority of scenarios considered, however, coverage probabilities are at reasonable levels.

      As outlined in Section~\ref{sec:nnhm}, instead of the overall mean~$\mu$, in some applications it may also make sense to consider estimation of one of the study-specific means~$\theta_i$ (shrinkage estimation), or of a ``new'' instance~$\theta_{k+1}$ (prediction). 
      For analogous results for prediction and shrinkage intervals, see Appendix~\ref{sec:PredShrinkAppendix}.
      In brief, one can see that the biases observed previously appear to propagate through, however, the coverage probabilities are substantially closer to their nominal levels.

\section{Applications}\label{sec:examples2}
  \subsection{Sorafenib example}
    \subsubsection{MTD estimation}
      Now consider MTD estimation in the context of the Sorafenib example (see Table~\ref{tab:sorafenibData}), again aiming for a DLT probability of $\pi^\star=0.33$. Considering the empirical DLT frequencies encountered, note that a number of studies never actually reach an empirical DLT frequency of 33\%, which also makes MTD estimation a bit of an extrapolation exercise here. Also, \emph{separation} occurs e.g. in the study by Furuse \emph{et~al.}\citep{FuruseEtAl2008}, where only two doses are investigated and events only were observed in one of the two groups. We will consider doses on their logarithmic scale for the analysis.
      Table~\ref{tab:sorafenibEstimates} (in Appendix~\ref{sec:ExampleDataAppendix}) also shows the resulting MTD estimates and their standard errors.
    
\begin{figure}[b]
        \centering
        \includegraphics[width=0.99\textwidth]{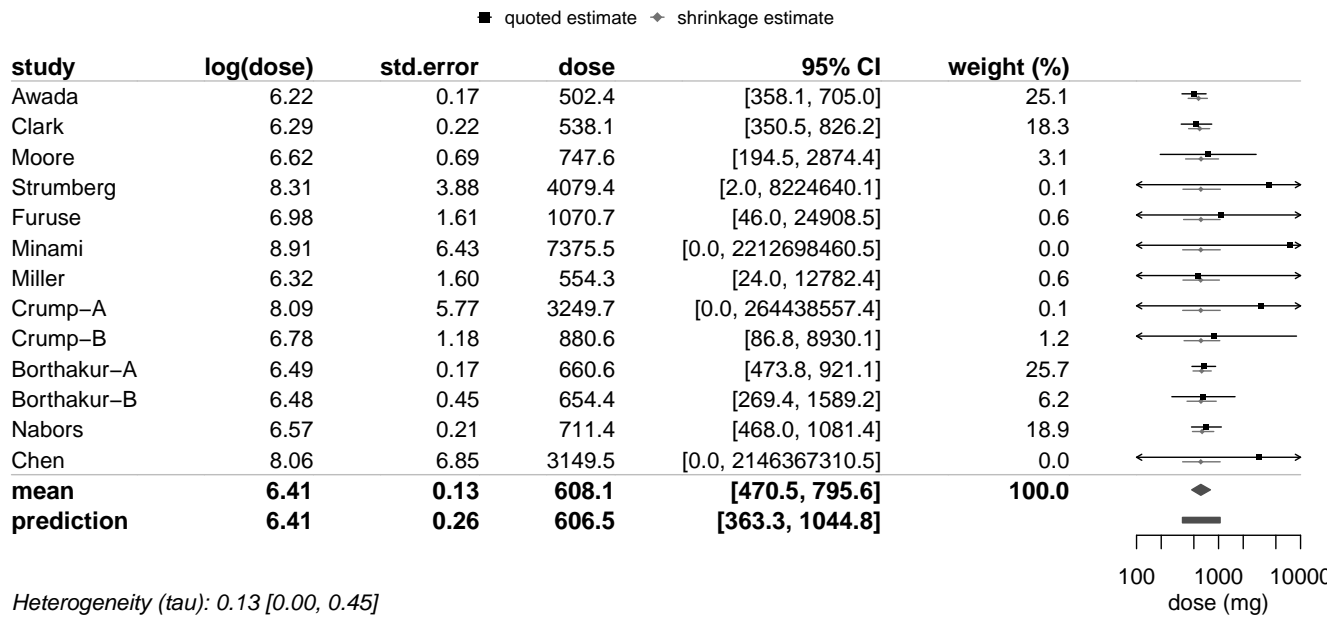}
        \caption{A forest plot illustrating the meta-analysis results for the Sorafenib example data set. MTD estimates are based on penalized logistic regression (FLAC). The estimates are shown in terms of logarithmic dose as well as on their original scale. The ``weight'' column indicates each study's contribution to the overall estimate.}
        \label{fig:sorafenibForest}
      \end{figure}

      The joint analysis of the 13~estimates is illustrated in Fig.~\ref{fig:sorafenibForest}.
      Note that the data from some of the studies barely allow to constrain the MTD, which is reflected in correspondingly large standard errors associated with the estimates. The eventual analysis is hence supported by certain studies, while others may not contribute substantially. The contribution of studies to the resulting estimate may be expressed in terms of (percentage) \emph{weights}, which are also shown in the Figure.\citep{RoeverFriede2021} 
      The eventual estimate of the overall mean MTD is at 608.1 [470.5, 795.6]. In terms of dose \emph{selection} (among the investigated doses; see Table~\ref{tab:sorafenibData}), this range only includes the dose of $600\,\mbox{mg}$. This is in agreement with the result previously found by Ursino \emph{et~al.},\citep{UrsinoEtAl2021} who worked using decrete dose levels. Therefore, the EMA-recommended dose of 400 mg (twice a day) is still considered a safer choice taking into account these new findings.\citep{EMA-Sorafenib}

      When the meta-analysis is performed at the design stage of a new study, it may be more relevant to look at a prediction interval; this may be useful to help designing the study, or to formally include the ``historical'' information in the eventual analysis. \citep{DeSantis2007,SchmidliEtAl2014,WandelNeuenschwanderRoeverFriede2017}
      The prediction interval is wider, as it includes the estimated between-study heterogeneity, and here it ranges from~$363.3$ to~$1044.8\,\mbox{mg}$.

      For any of the 13 included studies, we may also investigate the study-specific MTDs~$\theta_i$ in the light of the combined data via shrinkage estimation; for example the shrinkage estimate for the most recent study by Chen \emph{et~al.},\citep{ChenEtAl2014} which contributed only little information (a rather vague estimate of 3149.5 [0.0, 2146367310.5]), is at 607.0 [364.6, 1046.8]. As expected, this is very close to the prediction, but also substantially shorter than the original interval.

      The heterogeneity~($\tau$) is estimated at~0.13 [0.00, 0.45], which seems to be at a reasonable level for a log-transformed MTD\@.\citep{SpiegelhalterEtAl,RoeverEtAl2021}

      Given the weights that are shown in Figure~\ref{fig:sorafenibForest}, an obvious question is to what extent the studies with ``small'' weights are actually relevant to the analysis. The weights are closely related to the standard errors ($\sigma_i$) associated with each estimate,\citep{RoeverFriede2021} and it seems sensible that those studies with very large uncertainties associated may only help very little in constraining the MTD\@.  If we simply omit those studies with $\sigma_i>1$ (on the logarithmic scale this means that lower and upper 95\% interval bound are more than a factor of~$50$ apart), we are left with 6~studies, and the overall analysis results remain almost the same; see Figure~\ref{fig:sorafenibSubsetForest} in the Appendix.
      We may also check back with the original data (see Tables~\ref{tab:sorafenibData} and~\ref{tab:sorafenibEstimates}) and check why these particular studies only seem to provide little evidence; these are all studies either involving only few DLT~events, or studies that (empirically) exhibited a gently-sloping response, so that MTD estimates are large and at the same time associated with correspondingly large uncertainties.

    \subsubsection{Meta-analytic approach to bridging}
      Suppose one was interested in quantifying the MTD in Japanese patients. Strictly speaking, only two studies are available that investigated this figure\citep{FuruseEtAl2008,MinamiEtAl2008} (see also Table~\ref{tab:irinotecanData}). A meta-analysis of this pair of studies is readily performed and only requires the additional specification of a proper heterogeneity prior --- unlike in the case of many studies (large~$k$), where an improper uniform heterogeneity prior may be considered appropriate, the case of only $k=2$ studies requires a proper prior here.\citep{FriedeRoeverWandelNeuenschwander2017b,Roever2020,RoeverEtAl2021} For example, one may argue that a half-normal prior with scale~$0.2$ may be appropriate here, implying that while one anticipates some differences in study-specific MTDs, these  are expected to vary around their common mean by a factor ranging mostly between~$\frac{2}{3}$ and~$\frac{3}{2}$.\citep[Tab.~3]{RoeverEtAl2021}
      However, due to the low precision that the two Japanese studies provide, the meta-analysis also remains rather inconclusive with an estimated MTD of $1200$mg and a CI ranging from $56$mg up to $26\,000$mg. With this, the estimate covers \emph{all} of the potential doses (from $100$ to $1000$mg) that had been considered in any of the experiments.
      
      While evidence from the remaining European / North American studies may not be \emph{directly} transferable to the Japanese context, it seems obvious that still these are of some relevance here.\citep{ICH1998,EMA2013}
      A way to formally consider the external data for estimating the Japanese MTD in a dynamic fashion works via \emph{shrinkage estimation}.\citep{RoeverFriede2020}
      Technically, this may be implemented via several meta-analyses. Two summary estimates resulting from analyzing the groups of Japanese and Western studies separately are then combined in a third meta-analysis. At this second stage, the focus is not on an ``overall mean'' parameter, but rather on an updated \emph{shrinkage} estimate (here: the Japanese MTD).\citep{RoeverFriede2020} The assumption being implemented here is that there is heterogeneity within the Japanese and non-Japanese studies, as well as between the two groups of studies, resulting in an additional hierarchical model stage, and a total of three variance parameters.\citep{MorrisEtAl2018}  The second-stage meta-analysis again requires a proper heterogeneity prior; as in the above example, we will use a half-normal prior with scale~$0.2$. The results are shown in Table~\ref{tab:SorafenibBridging}.
      Excluding the two Japanese studies from the analysis yields an MTD estimate for the ``Western'' studies similar to the overall result shown in Table~\ref{fig:sorafenibForest}. The two Japanese studies alone on the other hand yield only a very vague estimate; the standard error (on the logarithmic scale) is roughly 10~times as wide as the ``Western'' standard error. Performing the meta-analysis at the second stage then yields a shrinkage estimate closer to the Western estimate and with a substantially reduced standard error.
      
      The new shrinkage estimate of 617 mg and the associated interval ranging from 337 up to 1179 mg do not allow to pinpoint a single dose, yet they allow to narrow down the range of likely and plausible doses considerably based on explicit and transparent model assumptions regarding the relationships between Japanese and external data. Again, we may also determine the \emph{weights} of the Western and Japanese studies as these contribute to the resulting shrinkage estimate;\citep{RoeverFriede2021} the Japanese contribution here amounts to only~$3.6\%$, once more highlighting the potential gain from considering the external data.

      \begin{table}[ht]
        \caption{Meta analysis aiming for an estimate of the MTD for a Japanese population. Assuming heterogeneity among Western and Japanese studies, as well as between the two groups of studies, first two separate analyses combining the Japanese studies\citep{FuruseEtAl2008,MinamiEtAl2008} as well as the remaining Western studies are performed. Subsequently the two resulting estimates are analyzed jointly, aiming for a shrinkage estimate of the Japanese MTD\@.}\label{tab:SorafenibBridging}
        \begin{center}
          \begin{tabular}{llcccc}
            \toprule
            & \multicolumn{2}{c}{log-MTD} & \multicolumn{2}{c}{MTD (mg)} \\
            \cmidrule(lr){2-3} \cmidrule(lr){4-5}
            analysis & mean & sd & median & 95\% CI\\
            \midrule
            Western (mean) & $6.41$ & $0.14$ & $606$ & [$467$, $794$] \\
            Japanese (mean)                  & $7.09$ & $1.57$ & $1199$ & [$56$, $25\,574$] \\[1.5ex]
            Japanese (shrinkage)             & $6.43$ & $0.30$ & $618$ & [$337$, $1179$] \\
            \bottomrule
          \end{tabular}
        \end{center}
      \end{table}

  \subsection{Irinotecan / S-1 example}
    Altough there is one study less in the second example, as well as fewer patients and fewer events involved, all studies saw an empirical DLT rate of at least 11\%, and there are eventually more studies that are able to contribute substantial information on the MTD (see also Table~\ref{tab:irinotecanData}). The resulting DLT estimates are shown in Table~\ref{tab:irinotecanEstimates} in Appendix~\ref{sec:ExampleDataAppendix}; only two studies yield (log-) DLT estimates with standard errors~$\sigma_i>1$. 
    One example of an ambiguous data setup is given by Yoshioka \emph{et~al.}\citep{YoshiokaEtAl2009}; the data do not even clearly suggest an increasing or decreasing toxicity with increasing dose (see Table~\ref{tab:irinotecanData}), and consequently the resulting MTD estimate ends up with a huge standard error. 
    Note also the noticeable differences between MTD estimates based on simple logistic regression compared to the regularized ones.

\begin{figure}
      \centering
      \includegraphics[width=0.99\textwidth]{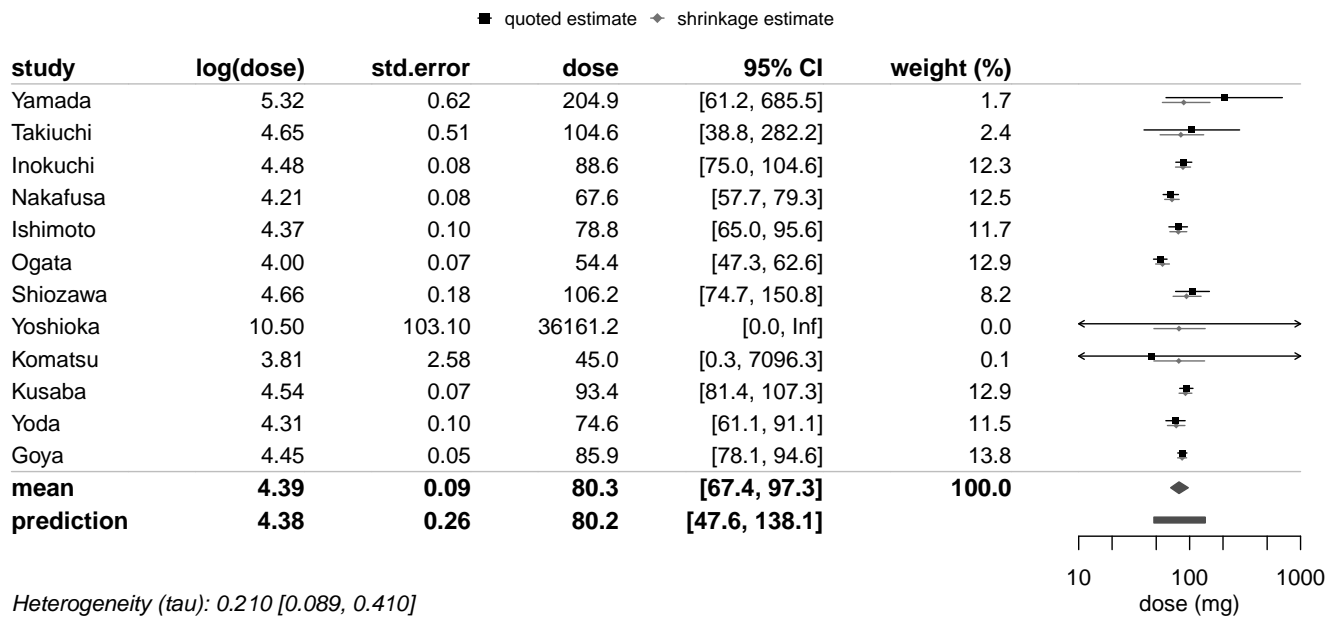}
      \caption{A forest plot illustrating the meta-analysis for the Irinotecan / \mbox{S-1} example data. MTD estimates are given on the logarithmic as well as on the dose scale.}
      \label{fig:irinotecanForest}
    \end{figure}

    Figure~\ref{fig:irinotecanForest} shows the results of a meta-analysis of all 12~MTD estimates.
    The eventual MTD estimate is at 80.3 [67.4, 97.3], which in terms of dose \emph{selection}, would include the doses of $70$, $80$ or $90\,\mbox{mg/m$^2$}$.
    To our knowledge, the combination of irinotecan and \mbox{S-1} has not yet received market authorization, but these figures are in agreement with the range determined by Ursino \emph{et~al.}\citep{UrsinoEtAl2021} for the MTD\@.
    The prediction interval, denoting the expected MTD in a new study, ranges from $47.6$ up to $138.1\,\mbox{mg/m$^2$}$. 

    We may again also consider shrinkage estimation; for example, for the most recent study (Goya \emph{et~al.}\citep{GoyaEtAl2012}), which contributed a pretty precise estimate of 85.9 [78.1, 94.6] on its own already, considering the remaining studies in addition does not add much information and the shrinkage estimate is very similar at 85.6 [77.9, 94.0].
    Another interesting example is the one by Yoshioka \emph{et~al.},\citep{YoshiokaEtAl2009} where one DLT was observed in 12~patients allocated to three doses ($100$, $125$ and $150\,\mathrm{mg}/\mathrm{m}^2$ ; see Table~\ref{tab:irinotecanData}).
    The authors eventually determined the highest dose as the recommended one, which seems appropriate considering only the data of the trial. We note, however, that the shrinkage interval for this study, $[47.6,\,138.1]$, excludes the dose of~$150$ (see also Figure~\ref{fig:irinotecanForest}).

\section{Discussion}\label{sec:conclusions}

 The aim of this manuscript is to propose a simple meta-analysis approach for the estimation of the maximum tolerated dose (MTD) from multiple phase~I dose-finding studies.
 The proposed  two-stage approach is easier to implement than a one-stage approach that would require additional assumptions on the model, such as the control group response or the implementation of pooling and stratification schemes.\citep{GuenhanEtAl2021}

 Based on our extensive simulation study, this simple approach gave on average good operating characteristis, in terms of MTD standard error (in the first stage) and MTD correct dose selection associated with DLT probability estimated at MTD (at the second stage). Only in the ``gentle'' scenario the proposed approach led to an underestimation of the selected MTD\@.  Unless the therapeutic window of the tested drug is very narrow (that is, if the minimum effective dose, MED, is close to the MTD), an underestimation of the MTD could be still considered acceptable, since a non-toxic, yet efficacious dose is selected. Neverthess, the ``gentle'' scenario reflects a case where an inappropriate dose range has been used in the original study.
 Indeed, if the toxicity probabilities associated with the investigated dose are similar, the first-stage regression (based on small samples) is likely to yield very imprecise MTD estimates, entailing problems with the delta method being applied to a non-linear response curve. The second-stage meta-analysis then may also not improve upon this. The selection of an appropriate set of doses in phase~I dose-finding trials is hence important in order to be able to derive useful MTD estimates.
 
 We also investigated the case of bridging studies where information from one population is used to extrapolate knowledge about another population. In the the Sorafenib illustration, we considered dynamic shrinkage estimation when using Western data for the estimation of the Japanese MTD\@. Ollier \emph{et~al}.~\cite{Ollier21} have proposed a complementary approach to evaluate the similarity between two population distributions of a parameter of interest (MTD, dose-response model parameter(s), etc.) and thus help in deciding whether or not to use and to tailor the available information.

 A limitation of this two-stage approach is related to the bias and coverage issues that seem to originate from the first-stage logistic regressions. A one-stage approach might avoid this, but would come at the cost of additional modeling assumptions and a more complicated analysis, as recalled at the beginning of this section.
 However, the heterogeneity parameter might `absorb' some of the erratic small-sample-size behaviour at the first stage, which might explain why  prediction and shrinkage estimates are not as badly affected.
 
 Bayesian approaches involve the specification of prior distributions, which is relevant for the meta-analysis at the second stage of our proposed approach. Here we relied on ``uninformative'' (uniform) priors for the overall mean and heterogeneity parameters as a default.
 In particular in cases involving few studies only, the use of (weakly) informative priors is recommended.\citep{Roever2020,RoeverEtAl2021}
 In practice, some prior information may come from earlier phases or from PK/PD considerations or preclinical data,\citep{ZhengEtAl2020} and it may be worth trying to include these in terms of informative priors.\citep{Roever2020,RoeverEtAl2021}
  
 Another point is associated with the way dose-finding designs are constructed. 
 In the interest of participants, early-phase dose-finding studies are designed with the aim to minimize the number of DLTs occuring, in particular when volunteers are involved. This is why these methods are often sequential, avoiding allocating patients to toxic doses and targeting the MTD\@.
 However, such schemes might result in large uncertainty in the eventual MTD estimate; from the estimation perspective, it may be more desirable to also investigate wider ranges of (not necessarily more toxic) doses in order to better probe the actual dose-response curve.\citep{Dette2008}
 
 Finally, our proposed simple meta-analysis method may be transferable to related estimation problems, e.g., for median lethal doses (LD\textsubscript{50}), estimation of the MED coming from phase~II dose-ranging studies, etc. User-friendly \textsf{R}~code is provided in the supplement to help clinical trials statisticians and stakeholders to implement the proposed method.

\vspace{10ex}

\section*{Data availability}
  The data that supports the findings of this study are available in the supplementary material of this article.


\begin{appendix}
\section{Example Data}\label{sec:ExampleDataAppendix}
  Tables~\ref{tab:sorafenibEstimates} and~\ref{tab:irinotecanEstimates} below characterize the data sets introduced in Section ~\ref{sec:examples1} in terms of the numbers of doses, patients and events, and show the slightly differing MTD estimates (on the logarithmic scale) from ``plain'' logistic regression, regression using Firth correction, and Firth's logistic regression with added covariate (FLAC) approaches.

  \begin{sidewaystable}[ht]
    \caption{Characteristics of the Sorafenib example introduced in Section~\ref{sec:examples1} (see also Table~\ref{tab:sorafenibData}); the numbers of doses, patients and DLT events are given for each study, as well as the MTD estimates (on the logarithmic scale) from plain logistic regression, regression using Firth correction and the FLAC approach.}\label{tab:sorafenibEstimates}
    \begin{center}
    \begin{tabular}{lrrrrrrrrrr}
      \toprule
      &&&&& \multicolumn{2}{c}{plain logistic} & \multicolumn{2}{c}{Firth} & \multicolumn{2}{c}{FLAC} \\
      \cmidrule(lr){6-7} \cmidrule(lr){8-9} \cmidrule(lr){10-11}
      publication                  & year & doses & patients & events & estimate & std. err. & estimate & std. err. & estimate & std. err. \\
      \midrule
      Awada \emph{et~al.} \citep{AwadaEtAl2005}        & 2005 &     6 &       37 &     10 & 6.22 &  0.15 & 6.19 &  0.17 & 6.22 & 0.17 \\
      Clark \emph{et~al.} \citep{ClarkEtAl2005}        & 2005 &     5 &       19 &      5 & 6.33 &  0.15 & 6.24 &  0.22 & 6.29 & 0.22 \\
      Moore \emph{et~al.} \citep{MooreEtAl2005}        & 2005 &     4 &       24 &      4 & 6.47 &  0.46 & 6.50 &  0.67 & 6.62 & 0.69 \\
      Strumberg \emph{et~al.} \citep{StrumbergEtAl2005}    & 2005 &     5 &       47 &      8 & 8.01 &  3.16 & 8.53 &  5.35 & 8.31 & 3.88 \\
      Furuse \emph{et~al.} \citep{FuruseEtAl2008}       & 2008 &     2 &       26 &      1 & 6.06 & 20.33 & 7.00 &  2.10 & 6.98 & 1.61 \\[1ex]
      Minami \emph{et~al.} \citep{MinamiEtAl2008}       & 2008 &     4 &       27 &      2 & 8.01 &  3.89 & 8.27 &  5.25 & 8.91 & 6.43 \\
      Miller \emph{et~al.} \citep{MillerEtAl2009}       & 2009 &     2 &       54 &     14 & 6.28 &  1.49 & 6.19 &  1.30 & 6.32 & 1.60 \\
      Crump \emph{et~al.} (A) \citep{CrumpEtAl2010}     & 2010 &     4 &       22 &      2 & 7.21 &  3.14 & 8.59 & 10.96 & 8.09 & 5.77 \\
      Crump \emph{et~al.} (B) \citep{CrumpEtAl2010}     & 2010 &     4 &       18 &      3 & 6.57 &  0.80 & 6.56 &  1.08 & 6.78 & 1.18 \\
      Borthakur \emph{et~al.} (A) \citep{BorthakurEtAl2011} & 2011 &     3 &       26 &      2 & 6.40 &  2.15 & 6.44 &  0.15 & 6.49 & 0.17 \\[1ex]
      Borthakur \emph{et~al.} (B) \citep{BorthakurEtAl2011} & 2011 &     3 &       16 &      3 & 6.37 &  0.25 & 6.38 &  0.43 & 6.48 & 0.45 \\
      Nabors \emph{et~al.} \citep{NaborsEtAl2011}       & 2011 &     5 &       20 &      5 & 6.57 &  0.17 & 6.52 &  0.21 & 6.57 & 0.21 \\
      Chen \emph{et~al.} \citep{ChenEtAl2014}         & 2014 &     2 &       19 &      1 & 6.07 & 30.28 & 3.10 & 13.91 & 8.06 & 6.85 \\
    \bottomrule
    \end{tabular}
    \end{center}
  \end{sidewaystable}

  \begin{sidewaystable}[ht]
    \caption{Characteristics of the Irinotecan / \mbox{S-1} example introduced in Section~\ref{sec:examples1} (see also Table~\ref{tab:irinotecanData}); the numbers of doses, patients and DLT events are given for each study, as well as the MTD estimates (on the logarithmic scale) from plain logistic regression, regression using Firth correction and the FLAC approach.}
    \label{tab:irinotecanEstimates}
    \begin{center}
    \begin{tabular}{lrrrrrrrrrr} 
      \toprule
      &&&&& \multicolumn{2}{c}{plain logistic} & \multicolumn{2}{c}{Firth} & \multicolumn{2}{c}{FLAC} \\
      \cmidrule(lr){6-7} \cmidrule(lr){8-9} \cmidrule(lr){10-11}
      publication              & year & doses & patients & events & estimate & std. err. & estimate & std. err. & estimate & std. err. \\
      \midrule
      Yamada \emph{et~al.} \citep{YamadaEtAl2003}   & 2003 &     3 &       12 &      1 & 5.02 &  2.63 &   5.28 &     0.81 &  5.32 &   0.62 \\
      Takiuchi \emph{et~al.} \citep{TakiuchiEtAl2005} & 2005 &     4 &       19 &      4 & 4.59 &  0.38 &   4.56 &     0.48 &  4.65 &   0.51 \\
      Inokuchi \emph{et~al.} \citep{InokuchiEtAl2006} & 2006 &     4 &       51 &     12 & 4.47 &  0.07 &   4.47 &     0.07 &  4.48 &   0.08 \\
      Nakafusa \emph{et~al.} \citep{NakafusaEtAl2008} & 2008 &     2 &       42 &      9 & 4.20 &  0.07 &   4.20 &     0.08 &  4.21 &   0.08 \\
      Ishimoto \emph{et~al.} \citep{IshimotoEtAl2009} & 2009 &     4 &       13 &      2 & 4.38 &  1.26 &   4.33 &     0.09 &  4.37 &   0.10 \\[1ex]
      Ogata \emph{et~al.} \citep{OgataEtAl2009}    & 2009 &     3 &       10 &      3 & 4.08 &  5.51 &   3.98 &     0.07 &  4.00 &   0.07 \\
      Shiozawa \emph{et~al.} \citep{ShiozawaEtAl2009} & 2009 &     4 &       21 &      7 & 4.67 &  0.16 &   4.64 &     0.18 &  4.66 &   0.18 \\
      Yoshioka \emph{et~al.} \citep{YoshiokaEtAl2009} & 2009 &     3 &       12 &      1 & 7.93 & 41.94 & -48.06 & 14805.91 & 10.50 & 103.10 \\
      Komatsu \emph{et~al.} \citep{KomatsuEtAl2010}  & 2010 &     3 &       21 &      2 & 4.07 &  1.56 &   3.86 &     3.02 &  3.81 &   2.58 \\
      Kusaba \emph{et~al.} \citep{KusabaEtAl2010}   & 2010 &     2 &        9 &      2 & 4.59 &  5.01 &   4.52 &     0.07 &  4.54 &   0.07 \\[1ex]
      Yoda \emph{et~al.} \citep{YodaEtAl2011}     & 2011 &     2 &        9 &      3 & 4.37 &  3.30 &   4.28 &     0.12 &  4.31 &   0.10 \\
      Goya \emph{et~al.} \citep{GoyaEtAl2012}     & 2012 &     3 &       11 &      3 & 4.49 &  1.65 &   4.44 &     0.05 &  4.45 &   0.05 \\
      \bottomrule
    \end{tabular}
    \end{center}
  \end{sidewaystable}

\clearpage

\section{Sensitivity analysis}
\begin{figure}[h]
        \centering
        \includegraphics[width=0.99\textwidth]{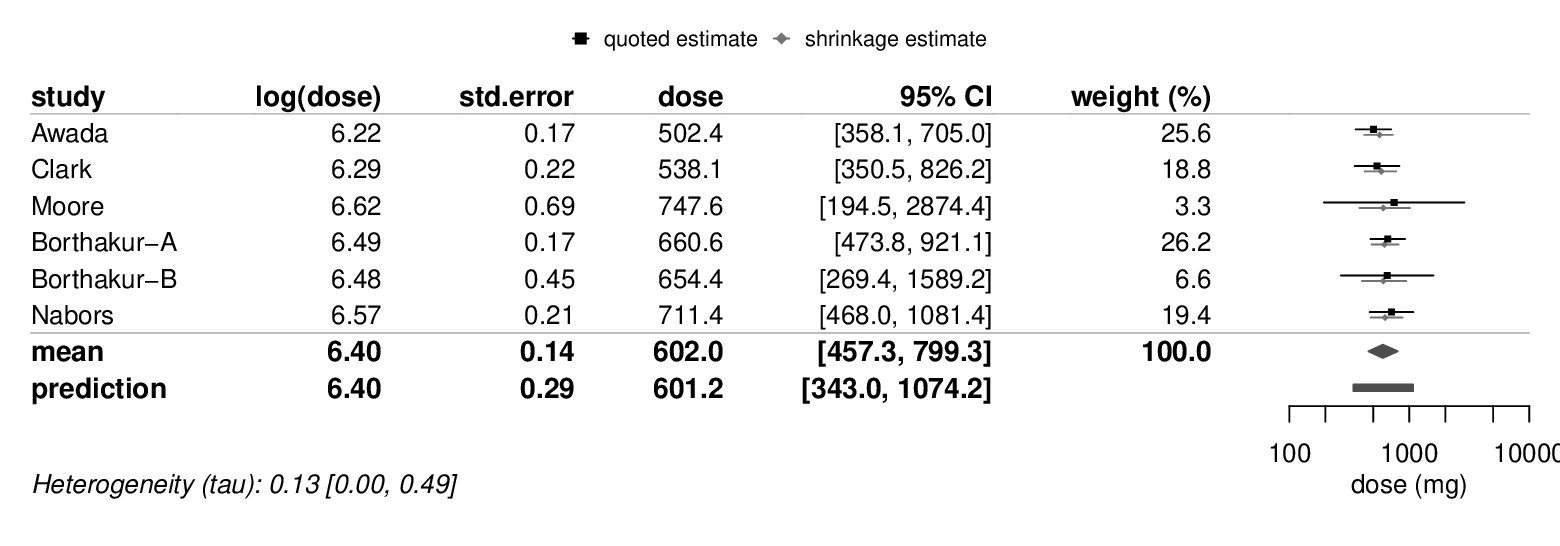}
        \caption{A sensitivity analysis for the sorafenib example data. MTD estimates are based on ``FLAC'' regressions. Omitting studies with ``large'' standard errors ($\sigma_i > 1$), which do not contribute much information on the overall mean estimate yields very similar results (see also Figure~\ref{fig:sorafenibForest}). }
        \label{fig:sorafenibSubsetForest}
      \end{figure}

  Figure~\ref{fig:sorafenibSubsetForest} illustrates a sensitivity analysis based on the Sorafenib data (see also Figure~\ref{fig:sorafenibForest}). A number of studies (those with ``large'' standard errors associated) contribute little to the analysis and hence have low weights assiciated. Omitting studies with low weights yields very similar, consistent results.

\section{Correspondence of logistic and Emax models}\label{sec:emaxAppendix}
  In case of a logarithmic dose as covariable ($x_j=\log(d_j)$, and
  $d_j>0$), the logistic model from Sec.~\ref{sec:methods} is
  equivalent to the \emph{Emax model}. The Emax model is defined as
  \begin{equation}\textstyle
    p_j \; = \; \emax(d_j) \; = \; E_\mathrm{max} \frac{d_j^n}{\Efifty^n + d^n}
    \mbox{,}
  \end{equation}
  where $\Emax>0$ denotes the maximal possible response, and
  $\Efifty>0$ denotes the dose level yielding half the maximum
  response.\citep{SchwinghammerKroboth1988}  In case of
  $\Emax\!=\!1$, the logistic and Emax models are equivalent; the
  Emax model's exponent~$n$ corresponds to the slope
  parameter~$\beta_1$, and the $\Efifty$~parameter is related to
  the intercept~$\beta_0$ as
  \begin{equation}
    \exp(-\beta_0) =  \Efifty^n
    \quad\Leftrightarrow\quad
    \beta_0 = -n \log(\Efifty)
    \quad\Leftrightarrow\quad
    \Efifty = \exp\Bigl(\textstyle - \frac{\beta_0}{n}\Bigr) 
    \mbox{.}
  \end{equation}

\section{Data generation details}\label{sec:DataGenerationAppendix}
  The simulation of data from common dose-finding designs  (see Section~\ref{sec:simul}) was facilitated by utilizing existing \textsf{R}~libraries.
  The ``\mbox{3+3}'' data were generated using the \texttt{UBCRM} package's ``\texttt{sim3p3()}'' function.\citep{R:ubcrm} The \mbox{3+3} simulation only requires specification of the (true) toxicity profile; the dose range is probed starting from the lower end, patients are recruited in cohorts of three, and decision on increasing, decreasing or proceeding with the current dose is done using fixed rules based on the number of DLTs observed in the previous cohort.\citep{ReinerPaolettiOQuigley1999,PaolettiEtAl2015}
  
  The ``CRM'' data generation was based on the \texttt{dfcrm} package's ``\texttt{crmsim()}'' function,\citep{R:dfcrm} using a cohort size of~3, the first dose as the starting dose, the skeleton as shown in Figure~\ref{fig:ExampleScenarios} and Table~\ref{tab:ExampleScenarios}, and a targeted DLT probability of $\pi^\star=0.33$. The default settings 	include a ``no-skipping rule''. The sample size was drawn uniformly (among multiples of~3) between~15 and~30. Patients are again recruited in small cohorts, and decision on the dose to be utilized for the next cohort is based on a parametric (logistic regression) model fitted to the past data.\citep{OQuigleyPepeFisher1990,GarrettMayer2006}
  
  ``BLRM'' data utilized the \texttt{bcrm} package's ``\texttt{bcrm()}'' function\citep{R:bcrm} using settings as above, and in addition using an EWOC criterion of~0.25 and vague independent lognormal priors for the logistic regression parameters (intercept, slope) with means~$\logit(0.01)$ and~$0.0$ and variances~$4$ and~$1$, respectively. The BLRM works similarly to the CRM, but is based on a Bayesian model and includes consideration of the (posterior) probability (here: 25\%) of exposing patients to overly toxic doses.\citep{NeuenschwanderEtAl2008,SweetingManderSabin2013,NeuenschwanderEtAl2015}

  Table~\ref{tab:ExampleDataProperties2} illustrates how the different designs employed in the simlations probe the dose-response curve in terms of the mean numbers of patients assigned to each of the six doses (see Section~\ref{sec:simul} for more details on the scenarios). Simulations here are based on ``fixed'' dose-response curves (no heterogeneity).
  \begin{table}[ht]
    \caption{Mean numbers of patients assigned to each dose in the different simulation scenarios, and based on the different experimental designs.}\label{tab:ExampleDataProperties2}
    \begin{center}
    \begin{tabular}{llcccccc}
      \toprule
      && \multicolumn{6}{c}{dose}  \\
      \cmidrule(lr){3-8}
      scenario                  & method & 1 & 2 & 3 & 4 & 5 & 6 \\
      \midrule
      \emph{moderate} & 3+3 & 3.15 & 3.35 & 3.79 & 3.69 & 1.82 & 0.27 \\
                      & CRM & 3.17 & 3.49 & 4.45 & 7.25 & 3.84 & 0.14 \\
                      & BLRM & 3.00 & 3.15 & 4.61 & 8.87 & 2.78 & 0.19 \\[1ex] 
      \emph{steep}    & 3+3 & 3.00 & 3.02 & 3.14 & 3.90 & 3.53 & 0.46 \\
                      & CRM & 3.00 & 3.03 & 3.17 & 5.23 & 7.75 & 0.28 \\
                      & BLRM & 3.00 & 3.00 & 3.05 & 7.48 & 5.74 & 0.33 \\[1ex] 
      \emph{gentle}   & 3+3 & 3.44 & 3.48 & 3.42 & 3.19 & 2.47 & 1.40 \\
                      & CRM & 3.50 & 3.78 & 4.39 & 5.69 & 4.63 & 0.65 \\
                      & BLRM & 3.03 & 3.22 & 4.22 & 6.19 & 4.26 & 1.83 \\[1ex] 
      \emph{convex}   & 3+3 & 3.13 & 3.21 & 3.33 & 3.89 & 3.26 & 0.43 \\
                      & CRM & 3.16 & 3.25 & 3.50 & 5.77 & 6.59 & 0.22 \\
                      & BLRM & 3.00 & 3.02 & 3.29 & 7.69 & 5.17 & 0.28 \\[1ex] 
      \emph{concave}  & 3+3 & 3.17 & 3.77 & 3.60 & 2.39 & 0.82 & 0.14 \\
                      & CRM & 3.18 & 4.58 & 6.96 & 6.43 & 1.33 & 0.05 \\
                      & BLRM & 3.05 & 4.03 & 7.54 & 6.39 & 1.33 & 0.09 \\
    \bottomrule
    \end{tabular}
    \end{center}
  \end{table}

\section{Prediction and shrinkage results}\label{sec:PredShrinkAppendix}
\begin{figure}
      \centering
      \includegraphics[width=0.85\textwidth]{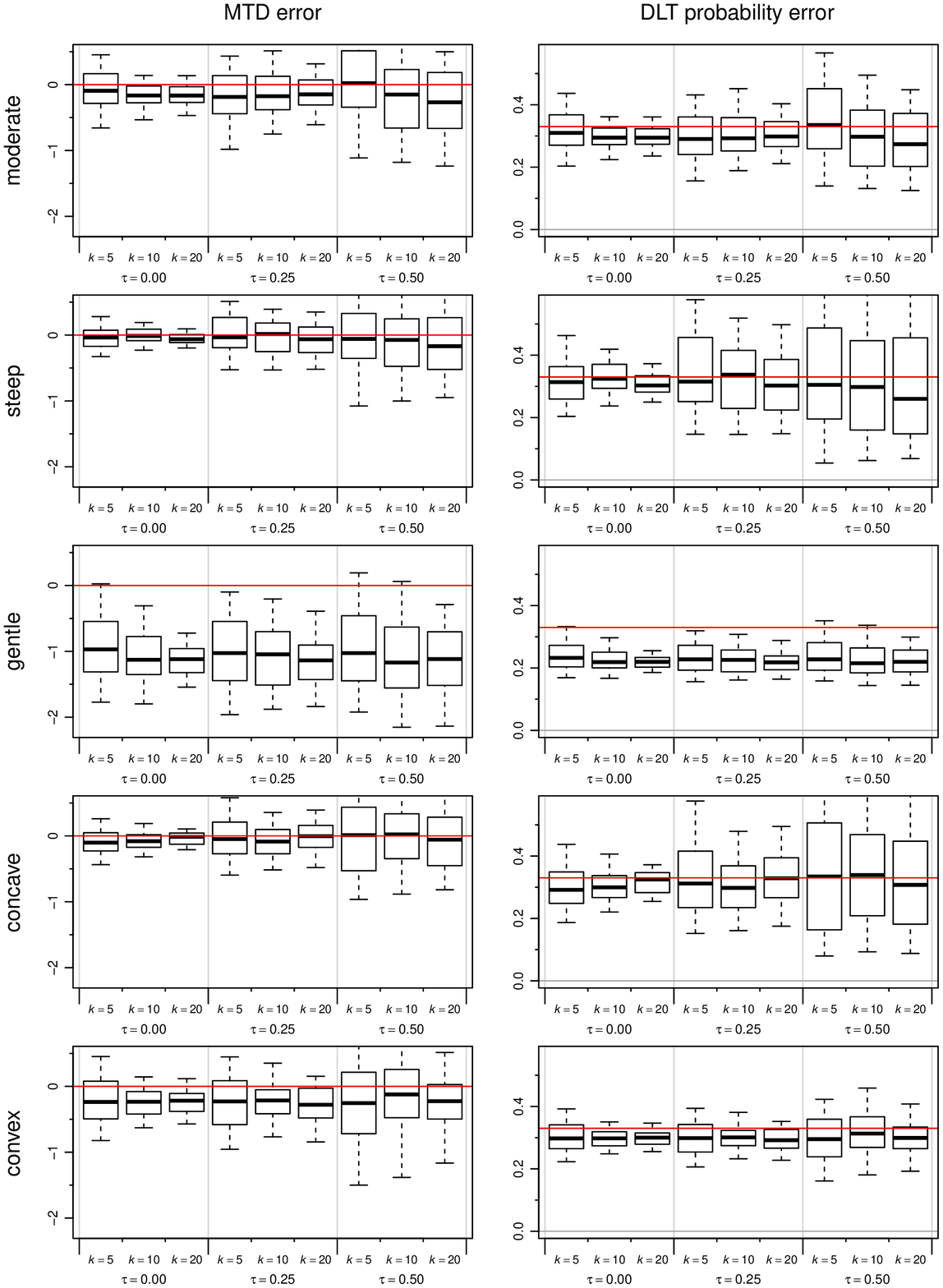}
      \caption{\textbf{Prediction} performance of meta-analysis based on FLAC estimates in different scenarios, and for different numbers of studies and amounts of heterogeneity. The left column shows the offset in estimated MTD in terms of \emph{doses}, and the right column shows the DLT probabilities at the estimated MTDs.
      (See also Fig.~\ref{fig:mtd-errors}.)}
      \label{fig:pred-errors}
    \end{figure}

\begin{figure}
      \centering
      \includegraphics[width=0.85\textwidth]{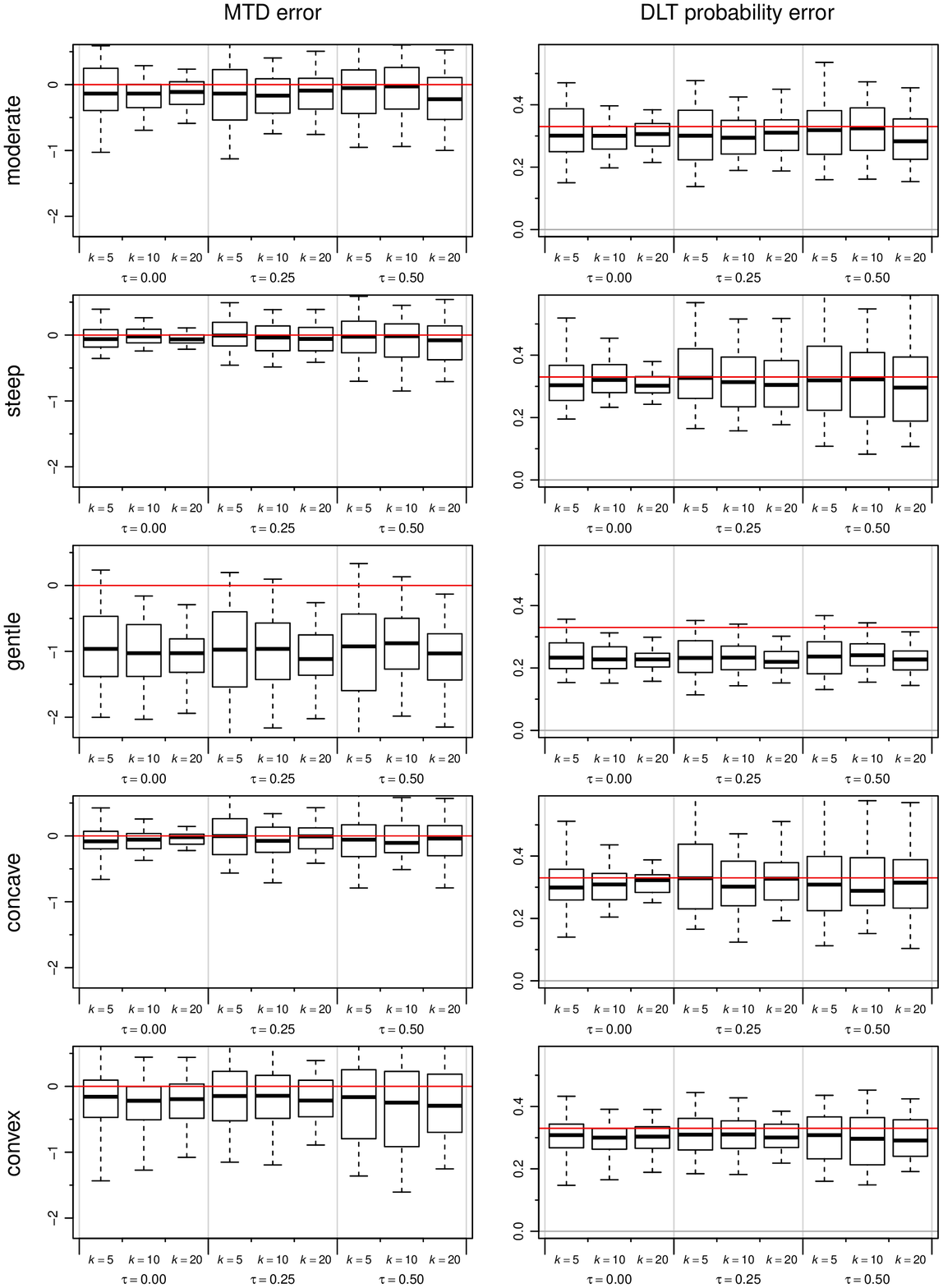}
      \caption{\textbf{Shrinkage} performance of meta-analysis based on FLAC estimates in different scenarios, and for different numbers of studies and amounts of heterogeneity. The left column shows the offset in estimated MTD in terms of \emph{doses}, and the right column shows the DLT probabilities at the estimated MTDs.
      (See also Fig.~\ref{fig:mtd-errors}.)}
      \label{fig:shrink-errors}
    \end{figure}

\begin{figure}
      \centering
      \includegraphics[width=0.85\textwidth]{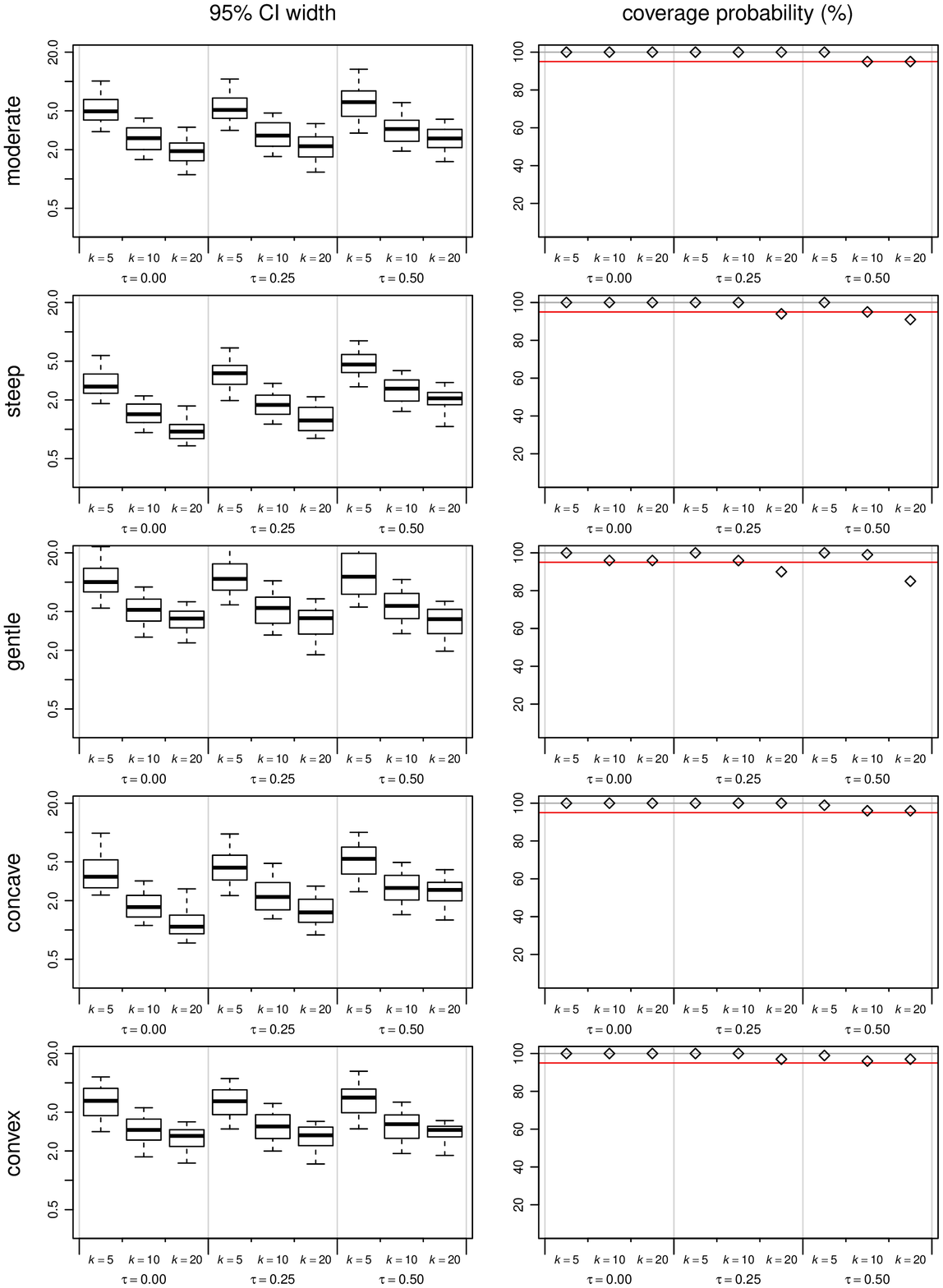}
      \caption{Widths and coverage probabilities of \textbf{prediction} intervals for the MTD based on the meta-analyses. (See also Fig.~\ref{fig:ci-mean})}
      \label{fig:ci-pred}
    \end{figure}

\begin{figure}
      \centering
      \includegraphics[width=0.85\textwidth]{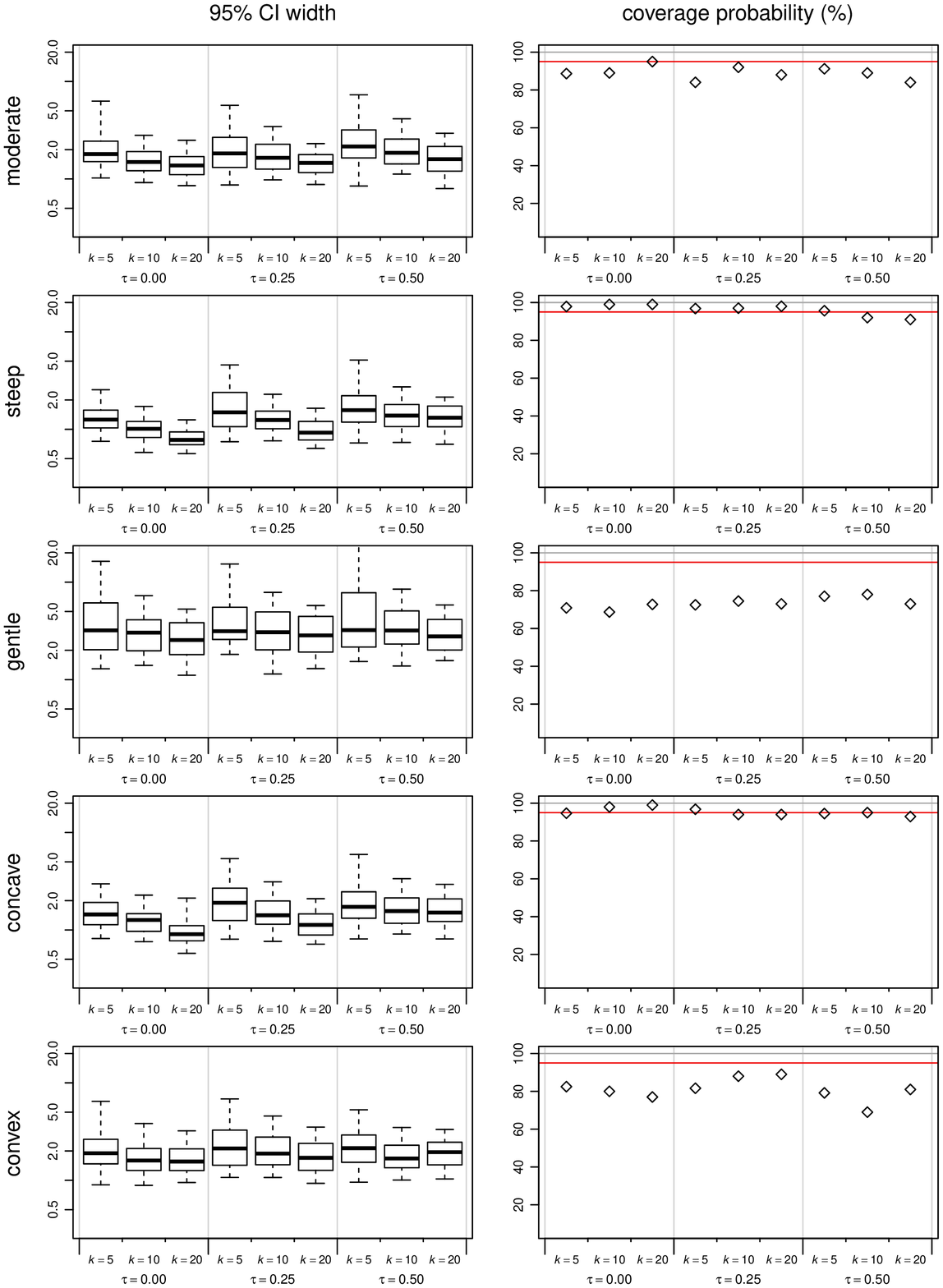}
      \caption{Widths and coverage probabilities of \textbf{shrinkage} credible intervals for the MTD based on the meta-analyses. (See also Fig.~\ref{fig:ci-mean})}
      \label{fig:ci-shrink}
    \end{figure}

  The four figures below illustrate the meta-analysis performance for prediction and shrinkage estimation.
  Figures~\ref{fig:pred-errors} and Figure~\ref{fig:shrink-errors} show the absolute offset in terms of \emph{dose levels} as well as \emph{DLT probability} in analogy to Figure~\ref{fig:mtd-errors}.
  Figures~\ref{fig:ci-pred} and~\ref{fig:ci-shrink} illustrate the credible interval lengths and their coverage probabilities analogously to Figure~\ref{fig:ci-mean}.

\end{appendix}

\clearpage

\bibliographystyle{wileyNJD-AMA}
\bibliography{literature,bilbio_MU} 

\end{document}